\documentclass[useAMS,usenatbib]{mnras}
\usepackage{newtxtext,newtxmath}
\usepackage[T1]{fontenc}
\usepackage{ae,aecompl}
\usepackage{graphicx}   % Including figure files
\usepackage{amsmath}    % Advanced maths commands
\usepackage{amssymb}    % Extra maths symbols
\usepackage{blkarray}
\usepackage{bigstrut}
\usepackage{layouts}
\usepackage[math]{cellspace}
\usepackage{breqn}
\usepackage{algorithm, algorithmic}

\usepackage{subcaption}
\usepackage{bm}
\usepackage[dvipsnames]{xcolor}
\numberwithin{equation}{section}

\newcommand{\zz}{\bmath{z}}
\newcommand{\zzc}{\bmath{z}^*} %\newcommand{\zzc}{\bmath{\bar z}}
\newcommand{\rr}{\bmath{r}}
\newcommand{\rrc}{\bmath{r}^*}%\newcommand{\rrc}{\bmath{\bar r}}
\newcommand{\vv}{\bmath{v}}
\newcommand{\vvc}{\bmath{v}^*} %\newcommand{\vvc}{\bmath{\bar v}}
\newcommand{\dd}{\bmath{d}}
\newcommand{\ddc}{\bmath{d}^*} %\newcommand{\ddc}{\bmath{\bar d}}
\newcommand{\Bmu}{\bm{\mu}}
\newcommand{\BLambda}{\bm{\Lambda}}

\newcommand{\AUG}[1]{\bmath{\breve{#1}}}
\newcommand{\Zz}{\AUG{z}}
\newcommand{\ZZ}{\AUG{Z}}
\newcommand{\Rr}{\AUG{r}}
\newcommand{\Dd}{\AUG{d}}
\newcommand{\RR}{\AUG{R}}
\newcommand{\Vv}{\AUG{v}}
\newcommand{\VV}{\AUG{V}}
\newcommand{\vDD}{\AUG{D}}
\newcommand{\augG}{\AUG{G}}
\newcommand{\vmV}{\vec{\mat{V}}}
\newcommand{\vmVH}{\vec{\mat{V}}^{*}}
\newcommand{\vmZ}{\vec{\mat{Z}}}
\newcommand{\vmZH}{\vec{\mat{Z}}^{*}}
\newcommand{\vmR}{\vec{\mat{R}}}
\newcommand{\vmRH}{\vec{\mat{R}}^{*}}

\newcommand{\G}{\mat{G}}
\newcommand{\GH}{\mat{G}^{H}}
\newcommand{\Q}{\mat{Q}}
\newcommand{\QH}{\mat{Q}^{H}}
\newcommand{\B}{\mat{B}}

\newcommand{\E}{\mat{E}}
\newcommand{\EH}{\mat{E}^{H}}
\newcommand{\K}{\mat{K}}
\newcommand{\KH}{\mat{K}^{H}}
\newcommand{\V}{\mat{V}}
\newcommand{\N}{\mat{N}}

\newcommand{\X}{\mathcal{X}}
\newcommand{\mat}[1]{{\mathbf{#1}}}
\newcommand{\JJ}{\mat{J}}
\newcommand{\DD}{\mat{D}}

\newcommand{\Bx}{\bm{x}}
\newcommand{\By}{\bm{y}}

\newcommand{\CN}{\mathrm{CN}} %{\mathcal{CN}}
\newcommand{\CST}[1]{\mathrm{CST}(#1)}
\newcommand{\Gam}[1]{\mathrm{Gam}(#1)}
\newcommand{\Matrix}[2]{\left [ \begin{array}{@{}#1@{}}#2\end{array} \right ]}

\newcommand{\Rop}[1]{\mathcal{R}_{{#1}}}
\newcommand{\Lop}[1]{\mathcal{L}_{{#1}}}

\definecolor{mygreen}{rgb}{0.0, 0.5, 0.0}
\newcommand{\noop}[1]{}
\everymath{\displaystyle}
\title[Robust-CubiCal]{Radio Interferometric Calibration Using a Complex Student's t-distribution and Wirtinger Derivatives}
\author[Sob et al.]{
U. M. Sob$^{1}$\thanks{E-mail: mulricharmel@gmail.com (UMS)},
H. L. Bester$^{2,1}$,
O. M. Smirnov$^{1,2}$,
J. S. Kenyon$^{1,2}$,
T. L. Grobler$^{3}$
\\
$^{1}$Department of Physics and Electronics, Rhodes University, Grahamstown, 6140, South Africa\\
$^{2}$South African Radio Astronomy Observatory, 2 Fir Street, Black River Park, Observatory, 7925, South Africa\\
$^{3}$Department of Mathematical Sciences, Computer Science Division, Stellenbosch University, Private Bag X1, 7602 Matieland, South Africa}
\date{Accepted XXX. Received YYY; in original form ZZZ}
\pubyear{2019}
\begin{document}
\label{firstpage} %https://www.overleaf.com/7878222212jsxfrnkcqptn
\pagerange{\pageref{firstpage}--\pageref{lastpage}}
\maketitle
% Abstract of the paper
\begin{abstract}
Radio interferometric gain calibration can be biased by incomplete sky models and radio frequency interference, resulting in calibration artefacts that can restrict the dynamic range of the resulting images. It has been suggested that calibration algorithms employing heavy-tailed likelihood functions are less susceptible to this due to their robustness against outliers in the data. We present an algorithm based on a Student's t-distribution which leverages the framework of complex optimisation and Wirtinger calculus for efficient and robust interferometric gain calibration. We integrate this algorithm as an option in the newly released calibration software package, CubiCal. We demonstrate that the algorithm can mitigate some of the biases introduced by incomplete sky models and radio frequency interference by applying it to both simulated and real data. Our results show significant improvements compared to a conventional least-squares solver which assumes a Gaussian likelihood function. Furthermore, we provide some insight into why the algorithm outperforms the conventional solver, and discuss specific scenarios (for both direction-independent and direction-dependent self-calibration) where this is expected to be the case. 
\end{abstract}
% Select between one and six entries from the list of approved keywords.
% Don't make up new ones.
%%%%%%%%%%%%%%%%%%%%%%%%%%%%%%%%%%%%%%%%%%%%%%%%%%
%%%%%%%%%%%%%%%%% BODY OF PAPER %%%%%%%%%%%%%%%%%%
\section{Introduction}
Calibration in radio interferometry is the data processing step during which unwanted propagation effects (e.g. ionospheric and tropospheric phase screens or complex antenna gain patterns) are estimated and removed from the data (known as visibilities). 
Since the visibilities and the various propagation effects are complex valued -- conventional optimisation theory deals with real functions of real arguments -- calibration is usually  performed by splitting these quantities into their real and imaginary parts\footnote{This transformation is motivated by the fact that calibration aims at minimising a real valued objective function of complex variables which is not holomorphic.}. 
Recent developments in the field of complex optimisation, particularly the application of Wirtinger calculus (see \citet{kreutzdelgado2009, sorber2012}), have made it possible to circumvent the need to split the data, and instead treat calibration as a complex optimisation problem.
As shown in \citet{tasse2014} and \citet{smirnov2015radio}, a complex formulation of radio interferometric (RI) calibration exploiting Wirtinger calculus bypasses the need to split the data and exposes properties which yield algorithmic advantages. 
Specifically, by careful ordering of the data, the Hessian of the optimisation problem can be adequately approximated by its diagonal, leading to significant algorithmic speed-ups.
\citet{grobler2018redundant} and \citet{kenyon2018cubical} show implementations of calibration algorithms based on these developments.

Since astrophysical signals of interest are typically very weak, the inevitable presence of radio frequency interference (RFI) introduces a significant challenge. 
RFI comes in many flavours, and it can be orders of magnitude brighter than the signals of interest, but (very fortunately for radio astronomy) it is often localised in time and frequency. 
The effects of strong RFI can usually be mitigated by rigorous data flagging (see \citet{offringa2012morphological} for example). 
The presence of weak or low-level RFI is actually more troublesome since it is more difficult to separate from the signal of interest. 
This is especially true for data processing pipelines which rely on automated data flagging algorithms that might require fine tuning to ensure that the signal has been accurately separated from the RFI. 
Sometimes, it is even necessary to resort to manual checking of the data. 
As a result, the removal of low-level RFI can quickly become a bottleneck for data processing pipelines. 
Fortunately, the impact of low level RFI on the calibration solutions can be reduced by using a suitable formulation of the calibration problem. 
From a statistical point of view, the presence of low level RFI is nothing more than outliers in the data, so its effects can be mitigated by using inference techniques which are robust against outliers.

Another potential source of bias is the incompleteness of sky models during calibration. 
Adopting the terminology of \citet{noordam2010meqtrees}, the process of first-generation calibration (1GC) solves for direction-independent effects (DIEs) using a ``known'' (relatively strong and isolated) calibrator source, and interpolates these solutions onto the visibilities of the target field. The resulting sky image is necessarily inaccurate (to a degree that is dependent on observing regime, instrumental stability, distance from the calibrator source, and/or the presence of direction-dependent effects, or DDEs), but it can usually be improved upon via the process of 2GC, or selfcal \citep{cornwell1981wilkinson}, which iterates between imaging and calibration, starting with the initial 1GC-derived model. This can be followed by 3GC, which solves for the DDEs. The limited accuracy of the initial 1GC-derived image, as well as limitations of deconvolution algorithms, means that only the brighter features of the initial image can be confidently taken to correspond to real emission, since lower-level features are increasingly likely to be contaminated by calibration artefacts. If such an image is used in a subsequent round of selfcal, these artefacts can be ``frozen'' into the resulting calibration solutions, and result in an non-physical map of the sky. Choosing the lesser of the two evils, we can start the 2GC process by including only the brighter sources in the calibration sky models, with subsequent iterations incorporating deeper models. However, the presence of fainter, unmodelled sources at each iteration can also result in calibration artefacts, which include the formation of ghost sources and suppression of real flux (see e.g. \citet{grobler2014calibration}, \citet{wijnholds2016calibration} and \citet{grobler2016calibration}). We note that while 2GC is particularly susceptible to this problem, 1GC can also be affected, since our models for the calibrator sources are not perfect and the newer telescopes may begin to resolve calibrator sources.

Mitigating these effects starts with understanding how the unmodelled sources affect the data or, more specifically, the residuals. As discussed in Section \ref{ssec:vist_dist}, a realistic unmodelled point source distribution results in a perceived increase in the overall noise level but doesn't alter the Gaussian shape profile of the noise distribution. The impact of unmodelled diffuse emission on the residuals is more subtle since it will affect different baselines in very different ways.  In particular, it is likely that the bias introduced by unmodelled diffuse sources depends strongly on the interplay between the solution intervals chosen during calibration and the power spectrum of the unmodelled diffuse flux. This will be investigated further in future work. In what follows we will restrict the discussion to unmodelled point sources and show that their impact on the calibration solutions can be minimised by using inference techniques which discourage over-fitting.   

Traditional calibration algorithms employ non-linear least squares (NLLS) algorithms such as the Levenberg-Marquardt (LM) or Gauss-Newton (GN) (see \citet{madsen2004methods}) which attempt to maximise a Gaussian likelihood function. However, since the presence of the outliers causes the data to deviate from the assumed Gaussian likelihood function, this is suboptimal. Such a straightforward maximum likelihood approach is also prone to over-fitting. Both these limitations have already been addressed in the literature (see \citet{kazemi2013robust} and \citet{ollier2017robust} for example) by using likelihood functions which are more robust against outliers and by imposing certain regularising constraints (e.g. smoothness in frequency). Most (if not all) of these approaches rely on splitting the data into its real and imaginary parts. {\color{black}We do note the existence of few non-GN/LM-based robust calibration algorithms. These include, for example, the first robust calibration algorithm by \citet{schwab1981vla} using an iterative relaxation method and $l_1$ regularisation, trust-region methods \citep{yatawatta2013radio} and quasi-Newton methods \citep{yatawatta2019stochastic}.}

In what follows we formulate the calibration problem as a complex valued iteratively re-weighted NLLS optimisation algorithm (the full details of which are given in Appendix~\ref{sec:appendix}). We illustrate that our approach automatically identifies and down-weights outliers in the data and is less prone to over-fitting (and hence flux suppression). Our implementation is based on the complex optimization formulation of \citet{tasse2014} and \citet{smirnov2015radio} and follows \citet{kazemi2013robust} in using a Student's t-distribution (ST) for the weight adaptation. The algorithm is made available as part of the newly developed calibration software package \emph{CubiCal} \citet{kenyon2018cubical} and can cope with fully polarised direction dependent calibration. 

The paper is organised as follows. In Section \ref{sec:calibration}, we provide a brief introduction to Wirtinger calculus and discuss how to apply it in the context of RI calibration. The details of the new algorithm are presented in Section \ref{sec:complex_t_cal}. Section \ref{sec:results} describes different simulations demonstrating how the implemented solver outperforms traditional solvers. These simulations investigate different regimes defined by the Signal to Noise Ratio (SNR) and the concentration of flux in the model visibilities. We also show that, in some cases, the implemented solver improves on the results from \citet{kazemi2013robust} because the covariance matrix of the residuals is estimated from the data and not assumed to be equal to the identity matrix. In Section \ref{sec:sim_rfi_data}, the algorithm is applied to synthetic and real data from the Karl G. Jansky Very Large Array (VLA) for low-level RFI mitigation. Possible extensions, as well as the limitations of the algorithm, are discussed in the conclusions.\\[0.1cm]

Notations: ${()}^T$, ${()}^*$ and $()^H$ denote the transpose, the complex conjugate and the complex conjugate transpose operators respectively. The $\mbox{vec}()$ operator converts a matrix to vector by stacking columns, $\mat{1}$ represents a matrix with all entries 1 while $\mathbf{I}$ is identity matrix. The superscript $\AUG{}$ denotes augmented vectors/matrices formed by stacking complex vectors/matrices with their conjugate counterparts. The over-arrow $\vec{}$ superscript is used to denote vectors of marices constructed by stacking matrices together.

\section{Calibration}\label{sec:calibration}
The Radio Interferometry Measurement Equation (RIME) \citep{hamaker1996understanding, smirnov2011revisiting} provides an elegant framework frequently used to model visibilities during calibration. In the RIME formalism, all distortions along the signal path are considered linear. Hence the model visibilities are represented by a series of matrix multiplications, each denoting a specific propagation effect. The RIME for a discrete source distribution\footnote{This form is only accurate for point sources. For extended sources it is usually assumed that the source is small enough so that we can approximate the DDE as being constant across the extent of the source.} can be written as
\begin{equation}
\qquad \quad \V_{pq} \,=\, \Q_{p} \,\left( \sum_{d}\E_{pd}\K_{pd}\B_{pqd}\KH_{qd} \EH_{qd}\right) \, \QH_{q}\, ,
\label{rime_1}
\end{equation}
where $\V_{pq}$ is the visibility measured between antennas $p$ and $q$, $\B_{pqd}$ is the assumed sky brightness matrix from the direction $d$, $\Q_{p}$ is the DI instrumental gain of antenna $p$, $\E_{pd}$ denotes the DD gain seen by antenna $p$ in direction $d$, and $\K_{pd}$ is the geometric phase delay associated with antenna $p$.

The measured visibilities and propagation effects in Equation \eqref{rime_1} are complex-valued. Usually, to circumvent taking complex derivatives\footnote{Since all real-valued functions of complex variables are not necessarily holomorphic.}, it is customary to split the data and the various propagation effects into their real and imaginary counterparts. In other words, the problem in $n$ complex-valued variables is replaced with a problem in $2n$ real-valued variables. Current developments in the field of complex optimisation allow bypassing the data transform (see for example \cite{kreutzdelgado2009} and \cite{sorber2012}) by using Wirtinger derivatives \citep{wirtinger1927}. In this section, we present the proposed calibration algorithm and the calibration suite in which it is implemented. A more extensive description of Wirtinger calculus and its applicability for RI calibration is available in \cite{smirnov2015radio} and \cite{kenyon2018cubical}.

\subsection{The Wirtinger approach}
\label{ssec:wirtinger}
Wirtinger calculus relies on treating the complex variable $\zz$ and its conjugate counterpart $\zzc$ as independent variables. Wirtinger derivatives are defined as  
\begin{equation}
\dfrac{\partial}{\partial z} = \dfrac{1}{2}\left ( \dfrac{\partial}{\partial x} - i\dfrac{\partial}{\partial y} \right),~~
\dfrac{\partial}{\partial z^*} = \dfrac{1}{2}\left ( \dfrac{\partial}{\partial x} + i\dfrac{\partial}{\partial y} \right),
\end{equation}
where $\zz = x+iy$, $\dfrac{\partial z}{\partial z^*} = 0$ and $\dfrac{\partial z^*}{\partial z} = 0$.
 
Consider the following optimisation problem
\begin{equation}
\label{eq:LSmin}
\min_{\bmath{z}} \,||\bmath{r}(\zz,\zzc)||_F \,=\, ||\bmath{d}-\bmath{v}(\zz,\zzc)||_F,
\end{equation}
where $\bmath{r}$, $\bmath{d}$, and $\bmath{v}$ are complex variables and $||.||_F$ is the Frobenius norm. This problem is solved by simply extending any NLLS algorithm such as the LM and the GN (see \cite{madsen2004methods}) to use Wirtinger derivatives. By treating $\zz$ and $\zzc$  as independent variables, we construct the following augmented vector for each of the unknown parameters
\begin{equation}
\label{eq:saugz}
\Zz = 
\begin{bmatrix}
\zz \\ \zzc
\end{bmatrix}.
\end{equation}
Furthermore, we augment all the functions to be functions of both variables $\zz$ and $\zzc$. Hence, we have the following for residuals, data and model respectively
\begin{equation}
\Rr = 
\begin{bmatrix}
\rr(\Zz) \\ \rrc(\Zz)
\end{bmatrix}, \; \Dd = 
\begin{bmatrix}
\dd(\Zz) \\ \ddc(\Zz)
\end{bmatrix},\; \Vv = 
\begin{bmatrix}
\vv(\Zz) \\ \vvc(\Zz)
\end{bmatrix}.
\end{equation}
Based on these definitions, the full Jacobian matrix, $\JJ$, is defined as
\begin{equation}
\label{eq:Jkrk}
\renewcommand{\arraystretch}{1.5}
\JJ = \dfrac{\partial \Vv}{\partial \Zz} = 
\cellspacetoplimit 3pt
\cellspacebottomlimit 3pt
\setlength{\arraycolsep}{4pt}
\begin{bmatrix}
\dfrac{\partial{\vv}}{\partial{\zz}} & \dfrac{\partial{\vv}}{\partial{\zzc}} \\ \dfrac{\partial{\vvc}}{\partial{\zz}} & \dfrac{\partial{\vvc}}{\partial{\zzc}}
\end{bmatrix}
= 
\Matrix{cc}{\JJ_{\vv\zz} & \JJ_{\vv\zzc} \\ \JJ_{\vvc\zz} & \JJ_{\vvc\zzc} }.
\end{equation}
The terms $\JJ_{\vv\zz}$, $\JJ_{\vv\zzc}$, $\JJ_{\vvc\zz}$ and $\JJ_{\vvc\zzc}$ are called partial and partial conjugate Jacobians. A deeper look shows that the diagonally adjacent terms are element-by-element conjugates of each other. From these definitions, the update steps for the parameters are defined as follows for the GN and LM algorithms
\begin{align}
GN; \qquad \delta\Zz &= (\JJ^H \JJ)^{-1}\JJ^H \Rr, \label{eq:GN} \\
LM; \qquad \delta\Zz &= (\JJ^H \JJ + \lambda\DD)^{-1}\JJ^H \Rr, \label{eq:LM}
\end{align}
where $\lambda$ is the LM damping factor and $\DD$ is the diagonalised Hessian matrix, $\JJ^H\JJ$.

Radio interferometers often employ antennas with different feeds to measure the polarisation properties of incoming radio waves. Therefore, the measured visibilities and propagation effects are generally not scalar complex variables but $2\times2$ complex variables or complex matrices. Fortunately, this does not make the optimisation problem intractable. All that is required is to vectorise the $2\times2$ complex matrices and derive the update steps given above. \cite{smirnov2015radio} (Appendix B) defines an operator calculus which makes the manipulation of $2\times2$ complex variables more convenient. For $2\times2$ complex variables, $\mat{Z}$, we define our augmented variables analogously to the scalar case i.e.
\begin{equation}
\label{eq:paugz}
\ZZ = 
\begin{bmatrix} 
\vmZ \\ \vmZH 
\end{bmatrix},
~~~\RR = 
\begin{bmatrix} 
\vmR(\ZZ) \\ \vmRH(\ZZ) 
\end{bmatrix},
~~~\VV = 
\begin{bmatrix}
\vmV(\ZZ) \\ \vmVH(\ZZ)
\end{bmatrix}.
\end{equation}
For example, $\vec{\mat{Z}}$ denotes the vector of matrices formed from all the parameters $\mat{Z}$. The quantities $\RR$ and $\VV$ are the augmented residuals and modelled visibilities respectively, expressed as functions of $2\times2$ complex matrices. The superscript ${}^*$ denotes element-wise complex conjugation. The full Jacobian matrix naturally follows as
\begin{equation}
\label{eq:pJkrk}
\JJ \,=\, \dfrac{\partial \VV}{\partial \ZZ} \, =\, 
\begin{bmatrix}
\dfrac{\partial{\vmV}}{\partial{\vmZ}} & \dfrac{\partial{\vmV}}{\partial{\vmZH}} \\ \dfrac{\partial{\vmVH}}{\partial{\vmZ}} & \dfrac{\partial{\vmVH}}{\partial{\vmZH}}
\end{bmatrix}.
\end{equation}
The derivatives that appear in Equation \eqref{eq:pJkrk} are matrix by matrix derivatives. These can be conveniently dealt with by using the operator calculus introduced by \cite{smirnov2015radio} which can be consulted for further details. The crucial result is that for any $2\times2$ matrices $\mat{A}$, $\mat{B}$ and $\mat{C}$, we have
\begin{equation}
\label{eq:pderiv}
\dfrac{\partial (\mat{ABC})}{\partial \mat{A}} = \Rop{\mat{C}}\Rop{\mat{B}},~~~\dfrac{\partial (\mat{ABC})}{\partial \mat{B}} = \Lop{\mat{A}}\Rop{\mat{C}},~~~\dfrac{\partial (\mat{ABC})}{\partial \mat{C}} = \Lop{\mat{A}}\Lop{\mat{B}},
\end{equation}
where $\Lop{\mat{A}}$ and $\Rop{\mat{A}}$ are matrix operators which act on $2\times2$ matrices called the left and right multipliers. They are defined such that for any $2\times2$ matrices $\mat{A}$ and $\mat{B}$:
\begin{equation}
\label{eq:pdefine}
\begin{array}{l@{~}l}
\Lop{\mat{A}}\mat{B} &= \mat{AB}, \\
\Rop{\mat{A}}\mat{B} &= \mat{BA}. 
\end{array}
\end{equation}
The key point here is that even for $2\times2$ complex variables, by carefully vectorising and using Eq. \eqref{eq:pderiv} and Eq. \eqref{eq:pdefine}, we end up with the following GN and LM update steps
\begin{align}
GN; \qquad \delta\ZZ &= (\JJ^H \JJ)^{-1}\JJ^H \RR, \label{eq:pGN} \\
LM; \qquad \delta\ZZ &= (\JJ^H \JJ + \lambda\DD)^{-1}\JJ^H \RR.\label{eq:pLM}
\end{align}
These are similar to those for the complex scalar case and can be implemented in an analogous way.  

\subsection{CubiCal overview}\label{ssec:cubical}
CubiCal \citep{kenyon2018cubical} is a recently developed software package which exploits complex optimisation. We provide a brief discussion of the software package, as our algorithm has been implemented as one of its subroutines.

A common bottleneck when implementing any NLLS algorithm is inverting the linearised approximation of the Hessian matrix, $\JJ^H \JJ$, appearing in \eqref{eq:pGN} and \eqref{eq:pLM}. \cite{smirnov2015radio} showed that, given a particular ordering of the solvable parameters (viz. antennas, directions, and correlations), this matrix is sparse in nature provided the problem is approached using Wirtinger calculus. Consequently it can be approximated by a diagonal matrix. CubiCal utilises this diagonal approximation to significantly reduce the computational cost of implementing the GN or LM update rules, albeit with slightly less accuracy. The algorithmic trade-off is that we usually require more of these significantly cheaper iterations to reach convergence. \citet{kenyon2019link} shows that this results in significant performance benefits in real-life cases.

CubiCal's modular structure makes implementing additional solvers, such as those presented here, relatively easy. In fact, all currently implemented CubiCal solvers (i.e. phase-only solvers, amplitude and phase solvers, parametrised slope solvers, and so on) could easily be augmented with Complex Student's t implementations. CubiCal read visibilities in the conventional Measurement Set data format. Model visibilities can be read from a measurement set or computed on-the-fly from a component sky model using the Montblanc package \citep{perkins2015}. This flexibility allows CubiCal to be incorporated into various 2GC and 3GC schemes with ease.

\section{Proper Complex Student's t Calibration}\label{sec:complex_t_cal}
This section details the implementation of the iteratively re-weighted complex NLLS solver (henceforth the \emph{robust solver}). In particular, we give the form of the proper complex Student's t-distribution (CST) as well as the update rules used for calibration. A full derivation is provided in Appendix~\ref{sec:appendix}. 

\subsection{Proper CST} 
\label{ssec:algo}
The Student's t-distribution (ST) is well known in the field of optimisation for its robustness in the presence of data containing outliers (see \cite{lange1989robust} for example), when compared with a Gaussian distribution. One way of constructing the ST is to visualise it as a mixture of random variables drawn from several Gaussian distributions with different standard deviations. We construct a CST by integrating a proper complex normal distribution over an unknown scale parameter $\tau$ for which we prescribe a Gamma prior i.e.

\begin{align}
\CST{\By| \Bmu, \BLambda, v} &= \int_0^\infty \CN\left(\By|\Bmu, (\tau \BLambda)^{-1}\right) \Gam{\tau| v, v} d \tau, \label{t-dist1}\\
  &= \int_0^\infty \frac{\tau^D |\bm{\Lambda}|}{\pi^D} \exp\left(-(\By - \bm{\mu})^H (\tau \BLambda) (\By - \bm{\mu}) \right) \nonumber\\
  &\qquad\times \frac{v^v \tau^{v-1} \exp(-v\tau)}{\Gamma(v)} d \tau 
\end{align}

\noindent where $\CN\left(\By|\Bmu, (\tau \BLambda)^{-1}\right)$ is a proper complex normal distribution with mean $\Bmu \in \mathbb{C}^D$ and Hermitian precision matrix $(\tau \BLambda) \in \mathbb{C}^{D\times D}$. $\Gam{\tau|v, v}$ is a Gamma distribution in $\tau \in \mathbb{R}^+$ and acts as a prior on the unknown scale parameter that we want to marginalise over. The resulting distribution takes the form
\begin{equation}
\CST{\By| \Bmu, \BLambda, v} = \frac{\Gamma(v+D) |\bm{\Lambda}|}{\Gamma(v) (v\pi)^D} \left(1 + \frac{\Delta^2}{v}\right)^{-v-D}. \label{marginal_dist}
\end{equation} 
Since this distribution is not a member of the exponential family, working with it directly is usually difficult. The standard way to overcome this difficulty is to utilise the Expectation Maximisation (EM) algorithm (see \cite{bishop2006pattern} for example). The EM algorithm iterates between estimating the expected value of the latent variables (missing or unavailable data) from their posterior distributions (E-step) and maximising the complete data likelihood function (M-step), which is generally easier to work with compared to the marginalised likelihood. For the full maximum likelihood solution, we need to solve for all the parameters of the CST distribution (viz. the means $\Bmu$, the precision matrix $\bm{\Lambda}$ and the number of degrees of freedom) during this step. 

For the specific problem of robust regression with a CST, the latent variables correspond to the scale parameter $\tau$ in Equation \eqref{t-dist1} whose posterior distribution is a Gamma distribution. 
As we show in Appendix~\ref{sec:appendix}, the solution can be obtained using an iteratively re-weighted complex NLLS algorithm in which the weights are computed as the expectation of $\tau$ under the Gamma posterior. 
An important aspect of the algorithm is that it adapts the likelihood used for calibration to the problem at hand by inferring the optimal number of degrees of freedom $v$. This parameter dictates the shape of the distribution. If the residuals are Gaussian, the inferred $v$ parameter will be large and we essentially recover a Gaussian likelihood. If the residuals contain outliers, the inferred $v$ parameter will be small and data points which could otherwise bias the optimisation procedure get down-weighted and therefore do not significantly affect the calibration solutions. Some further insight into this behaviour is provided below. 

\subsection{Robust Calibration}
Consider the following form of the RIME for a field with $N_d$ sources
\begin{equation}
 \V_{pqs} \,=\, \sum_{d=1}^{N_d}{\G_{ps}(d) \X_{pqs}(d) \GH_{qs}(d)} + \epsilon_{pqs},  \label{t-rime}
\end{equation}
where $G_{ps}$ is the gain for antenna $p$, $\mathcal{X}_{pqs}$ is the sky coherency in direction $d$, $s$ is the corresponding time and frequency index and $\epsilon_{pqs}$ is the noise which is assumed to be CST distributed. Thus calibration can be performed as described in Appendix~\ref{sec:appendix}. Given initial values for the weights, $\mat{W}$, the Jones matrices, or gains, can be computed by minimising the following objective function
\begin{equation}
\label{eq:wLSmin}
\min_{\G} \;||\mat{W}\left(\RR(\G,\GH)\right)||_F \;=\; \min_{G}\;||\mat{W}\left(\vDD - \VV (\G,\GH)\right)||_F,
\end{equation}
where $\G$ is the gain matrix and $\RR$, $\vDD$ and $\VV$ are the augmented residual, data and model vectors respectively. The elements of the $\mat{W}$ matrix are updated at each iteration and given by the the expectation values of the latent variables $\tau_i$ of the CST. Explicitly, they can be written as
\begin{equation}
\mat{w}_{pqs} \; =\; \frac{v+n_c}{v + \mat{R}_{pqs}^{^H} {\bf{\Sigma}}^{-1}\mat{R}_{pqs}} \label{gain_w},
\end{equation}
where $\mat{w}_{pqs}$ represents the weight of the $2\times2$ visibility matrix between antenna $p$ and $q$ at time and frequency index $s$, $n_c$ is the number of correlations in our data and $\mat{R}_{pqs} = \mbox{vec}(\RR_{pqs})$ is the residual of the corresponding visibility.  Note that $\mat{R}_{pqs}$ here is a $4\times1$ vector and not a $2\times2$ matrix, as expected from the vectorisation. ${\bf{\Sigma}}$ is the covariance matrix of the residual visibilities and it is a $4\times4$ matrix we generally assume to be diagonal. The number of correlations, $n_c$, is important because, even though CubiCal assumes a data structure where each visibility is a $2\times2$ matrix, for scalar calibration or data with single correlations, the cross correlation terms are set to zero. Hence, $n_c$, which represents the dimension of a single vectorised visibility, is effectively 4 only when all the correlations are present. Note that the $\mat{R}_{pqs}^{^H}{\bf{\Sigma}}^{-1}\mat{R}_{pqs}$ term in the denominator will have an expectation value of $n_c$ if the data are Gaussian distributed with covariance matrix ${\bf{\Sigma}}$. 

The $v$-term is computed by solving the following equation
\begin{dmath}
-\psi(v) + log(v) + 1 + \psi(v+n_c) - log(v+n_c) 
+ \frac{1}{N}\sum_{pq}(log(\mat{w}_{pqs}) - \mat{w}_{pqs}) = 0, \label{v_equation}
\end{dmath}
where $\psi$ is called the digamma function (logarithmic derivative of the gamma function) and $N$ is the total number of visibilities. Equation \eqref{v_equation} has no closed form solution and has to be solved numerically. We find that, in practice, it is sufficient to restrict $v$ to be an integer and to simply do a grid search between $2\leq v \leq 50$ since, as already mentioned, the ST is almost indistinguishable from a Gaussian when $v>30$ or so. Finally, at each iteration, the covariance matrix ${\bf{\Sigma}}$ is computed as follows
\begin{equation}
\label{eq:sigma}
{\bf{\Sigma}} = \frac{1}{N}\sum_{pqs}\left(\mat{R}_{pqs}\mat{R}_{pqs}^{H}\mat{w}_{pqs}\right),
\end{equation}
where $N$ is again the total number of visibilities.

A closer look at Equation \eqref{gain_w} can provide some insight into the workings of the robust solver. Clearly, the solver assigns small weights to visibilities with large residuals and large weights to visibilities with small residuals\footnote{Note the upper bound on the weights is finite and equal to $\frac{v + n_c}{v}$.}. When the residuals follow a Gaussian distribution with covariance ${\bf{\Sigma}}$, the $v$-term is large and all the visibilities end up having approximately equal weights. On the other hand, for visibilities containing outliers, the $v$-term is small and the outliers can be effectively down-weighted. Finally, suppose that the covariance has been under-estimated (as will be the case if the residuals also contain a realistic unmodelled point source distribution). In this case, the $\mat{R}_{pqs}^{^H} {\bf{\Sigma}}^{-1}\mat{R}_{pqs}$ term in the denominator will be much larger than $n_c$ and these points will be down-weighted, thus discouraging over-fitting.

\subsection{Implementation details}
\label{ssec:impl}
Algorithm \ref{algo1}  shows the details of the new algorithm which we have dubbed the robust solver. The robust solver implementation was greatly simplified thanks to CubiCal's object-oriented programming approach. CubiCal provides an abstract class interface with preset attributes and functions which need to be inherited and defined to develop any new solver. In CubiCal terminology, we refer to this as a Gain Machine. The new solver is invoked in CubiCal by setting the solver's option \emph{gain-type} to \emph{robust-2x2}.
 \begin{algorithm} % enter the algorithm environment
\caption{: Robust Solver Algorithm} % give the algorithm a caption
\label{algo1} % and a label for \ref{} commands later in the document
\begin{algorithmic} % enter the algorithmic environment
    \REQUIRE {$\text{Data}\; \vDD, \text{Model}\; \VV, \;\text{Jacobian_func},\; i_{\mathrm{max}}$}
    \STATE {$\textbf{Initialisation:\,} \augG_{0} \gets \mat{1},\, \mat{w}_{pqs} \gets 1,\, v \gets 2, i \gets 0$}    
    \WHILE{(not \textbf{converged} or not \textbf{stalled} or $i \leq i_{\mathrm{max}}$)}
         \STATE {$\mat{W} \, \gets \, Diag(\mat{w}_{pqs})\;$} \COMMENT{\# Diagonal matrix with weights}\\
         \STATE {$\JJ \gets \text{Jacobian_func\ } (\vDD, \VV, \augG_{i-1})$}
         \STATE {$\RR \gets \vDD  - \VV$}
         \STATE {${\bf{\Sigma}} \gets $ Update ${\bf{\Sigma}}$ using $\RR$, $\mat{W}$ and Eq. \eqref{eq:sigma}}
         \STATE {$\delta \augG  \gets  (\JJ^H\mat{W}\JJ)^{-1}\JJ^H\mat{W}\RR$\\}
         \STATE {$\augG_{\mathrm{temp}} \gets \augG_{i-1} + \delta \augG$\\}
        \IF{ i mod 2 = \,0 or DD calibration\\}
            \STATE $\augG_{i} \, \gets \frac{1}{2} \left(\augG_{\mathrm{temp}} + \augG_{i-1}\right)$
        \ELSE
            \STATE $\augG_{i} \, \gets  \augG_{\mathrm{temp}}$
        \ENDIF
        \FOR {all baselines}
            \STATE $\mat{w}_{pqs}\, \gets \frac{v+n_c}{v + \mat{R}_{pqs}^{^H} {\bf{\Sigma}}^{-1}\mat{R}_{pqs}} $
        \ENDFOR
        % \FOR {all baselines pairs $pq$}
        %   \STATE $\mat{w}_{pqs} \gets \frac{\mat{w}_{pqs}}{Average(\mat{w}_{pqs})}\;$ \COMMENT{\# Scale the weights}
        % \ENDFOR
        \STATE {$v\, \gets$ Compute $v$ using Eq. \eqref{v_equation} \\}
        \STATE {$i \, \gets\, i +1$}
    \ENDWHILE
\end{algorithmic}
\end{algorithm}

The expected thermal noise level for the observation is used to pre-whiten the data. This means that the weights can be initialised to $\mat{1}$ during the first iteration. We treat them as scalar real variables meaning all correlations have the same weight. Furthermore, the weights are assigned per visibility, independently of time, frequency or baseline. As is customary in radio interferometry, the weights of flagged data are set to zero from the start. The computation of the weights involves the residual covariance matrix, ${\bf{\Sigma}}$, which is not included in \cite{kazemi2013robust} but just assumed to be $\mathbf{I}$. We do not make this assumption. Instead, we implement two variants of the algorithm, one with ${\bf{\Sigma}}$ computed using \eqref{eq:sigma}, and another where we set ${\bf{\Sigma}}$ to $\mathbf{I}$. A setting is made available to the user to decide whether or not they want ${\bf{\Sigma}}$ to be computed during every iteration or simply set it to $\mathbf{I}$. The default behaviour of the solver is to compute ${\bf{\Sigma}}$ as it is more consistent with our derivation of the algorithm (see Appendix~\ref{sec:appendix}). Furthermore, we also provide an option to fix the number of degrees of freedom at the outset without inferring it using \eqref{v_equation}.

It has been observed that averaging the gain solutions every second iteration improves the convergence speed of the algorithm (see \citet{salvini2014fast}). \citet{smirnov2015radio} explains that this averaging corresponds to alternating between the GN and LM algorithms. This is very helpful when calibrating for DD effects as these generally converge slowly. For the CubiCal solver, we average solutions at every iteration for DD calibration, and at even iterations for DI calibration.

\subsection{Computational cost}
\label{ssec:cost}
The main additional operations performed by the robust solver are the computations of the weights and the numerical solution for the degrees of freedom, $v$. Assigning the weights relies on computing the residual visibilities and the covariance matrix ${\bf{\Sigma}}$. The algorithm is implemented such that the residuals are computed only once during every iteration. The residuals computed for the weight updates are stored in memory and reused during the gain updates. For DI calibration, the default solver does not compute residual visibilities at every iteration. This is made possible thanks to an observation from \cite{tasse2014}. For DI calibration, we have the following RIME form
\begin{equation}
\VV = \AUG{G} \AUG{M} \AUG{G}^{H},  \label{DImodel}
\end{equation} 
where $\AUG{M}$ corresponds to the true or modelled visibilities. \cite{tasse2014} states, 
\begin{equation}
\label{eq:jtrick}
\VV = \JJ_{\mathrm{L}} \G = \frac{1}{2} \JJ \AUG{G},
\end{equation}
where $(\cdot)_{\mathrm{L}}$ denotes the left half of a matrix and $\AUG{G} = 
\begin{bmatrix} 
\G \\ \G^{H} 
\end{bmatrix}$. Substituting Equation \eqref{eq:jtrick} in Equation \eqref{eq:pGN}, we have the update rule for DI calibration below
\begin{align}
\label{eq:subr}
\delta\G &= (\JJ^H \JJ)^{-1}_{\mathrm{U}} \JJ^H (\AUG{D} - \JJ_{\mathrm{L}} \G)\\
 &= (\JJ^H \JJ)^{-1}_{\mathrm{U}} \JJ^H \AUG{D} - \G,  
\end{align}
where $(\cdot)_{\mathrm{U}}$ stands for the upper half of a matrix. This implies that
\begin{align}
\label{eq:updG}
\G_{i} &= \G_{i-1} + \delta\G \\
 &= (\JJ^H \JJ)^{-1}_{\mathrm{U}} \JJ^H \AUG{D}.
\end{align}
Hence, residuals are not required for updating the gains.
In the case of DD calibration Equation \eqref{eq:jtrick} does not hold, and both solvers have to compute residuals at each iteration. Fortunately, CubiCal employs various levels of parallelism, and we script the most expensive tasks in the Cython programming language (note that in the latest version Cython has been replaced with Numba). These dramatically improve the speed for generating the necessary residual visibilities. Additionally, in CubiCal only the diagonal of the Hessian is computed and the full Jacobian matrix is never loaded into memory but is instead implemented as an operator.

CubiCal uses the below data structure
\begin{equation}
\mat{\AUG{D}} = 
\begingroup
\begin{bmatrix}
    \mat{0} & \mat{D}_{12} & \mat{D}_{13} &\dots& \mat{D}_{1N_a}\\
    \mat{D}^H_{12} & \mat{0} & \mat{D}_{23}&\dots& \mat{D}_{2N_a}\\
    \mat{D}^H_{13} & \mat{D}^H_{23} & \mat{0}&\dots& \mat{D}_{3N_a}\\
     \vdots & \ddots & \ddots &\ddots &\ddots\\
     \mat{D}^H_{1N_a}& \mat{D}^H_{2N_a} & \mat{D}^H_{3N_a}&\dots&\mat{0}
\end{bmatrix},
\endgroup
\end{equation}
where $N_a$ is the number of antennas and each element is a $2\times2$ complex matrix. Half of the data is just the conjugate transpose of the other half. This implies only half of the data is required to compute the covariance matrix ${\bf{\Sigma}}$. Similarly, half of the weights are sufficient to solve for $v$.
Another optimisation strategy is to update $v$ only after a specific number iterations. The number of iterations after which to recompute $v$ is a setting which can be modified by the user. Moreover, we restrict the search space for $v$ by assuming it is an integer and performing a grid search between 2 and 50\footnote{We do this by computing the function at different $v$ integer positions and take the $v$ position with the minimum value as the solution. We avoid using numerical solvers as they may introduce convergence issues or slow the solver since we only need an estimate of this value.}.

\section{Robust solvers and flux suppression}
\label{sec:results}
This section uses simulated data to identify some regimes in which the robust solver can be expected to improve the results of calibration. Our main aim is to compare how much of the unmodelled flux gets suppressed during calibration with the different solvers. For brevity, we refer to them as follows:

\newcommand{\robustI}{robust-$\mathbf{I}$}

\begin{itemize}
\item ``complex solver'': a conventional least-squares solver employing the Wirtinger formulation (identified as ``cp'' in figure legends).

\item ``robust solver'' with covariance iteratively recomputed (identified as ``rb'' in figure legends).

\item ``robust-$\mathbf{I}$ solver'' for the robust solver with covariance set to $\mathbf{I}$ (identified as ``rb-$\mathbf{I}$'' in figure legends).
\end{itemize}
To aid our understanding of when the robust solver can be expected to out-perform the traditional solver, we start with a simple illustration of how unmodelled sources affect the statistics of the residual visibilities.

\subsection{Statistical properties of visibilities}
\label{ssec:vist_dist}
\begin{figure*}
\begin{subfigure}{.48\textwidth}
  \includegraphics[width = 0.95\linewidth]{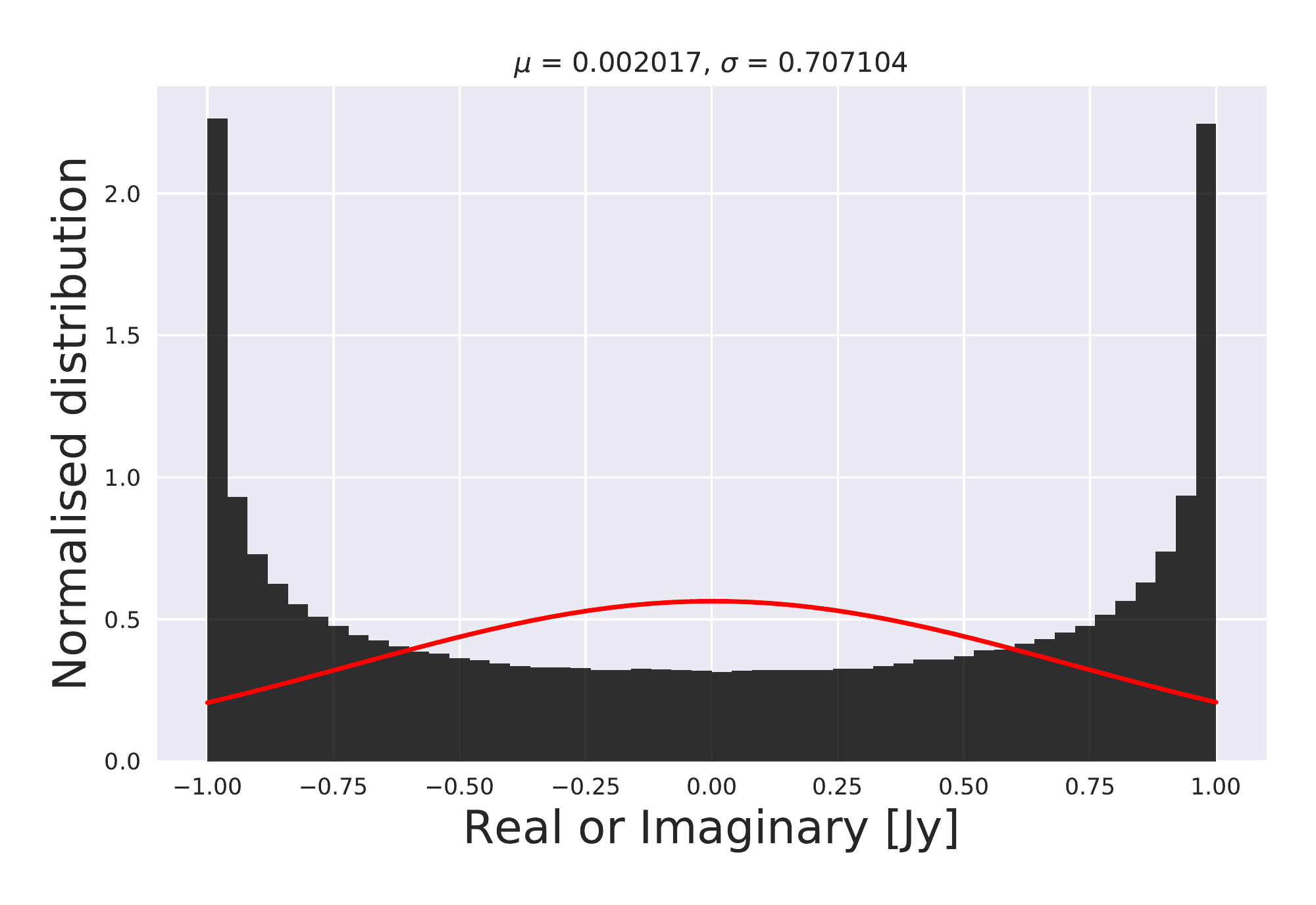}
  \caption{1 Jy source at an offset position }\label{hist1}
 \end{subfigure}
\begin{subfigure}{.48\textwidth}
  \includegraphics[width = 0.95\linewidth]{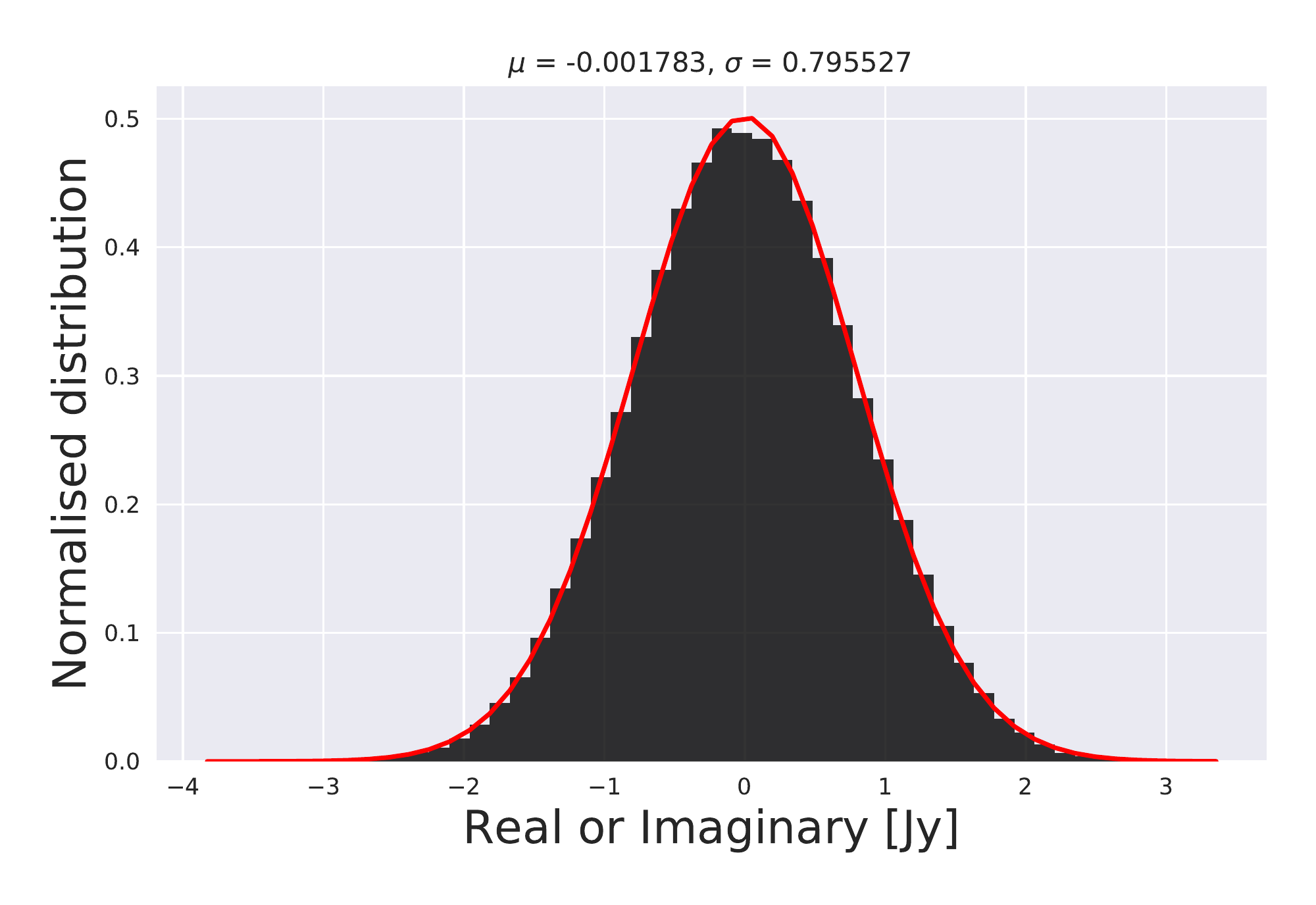} 
\caption{100 sources}\label{hist2}
 \end{subfigure}
\caption{(a): {\color{black}Histogram of simulated visibilities of a 1 Jy source at an offset position from the phase centre of the field.} (b): Histogram of simulated visibilities for 100 sources drawn from a realistic sky model.  The red curve in each plot is the corresponding Gaussian probability function computed using the mean and standard deviation of the visibilities. The mean, $\mu$ and standard deviation, ${\bf{\Sigma}}$ are shown respectively on the figure titles.}
\label{hist}
\end{figure*}

Calibration with incomplete sky models implies that the residuals which we attempt to minimise during the optimisation process (calibration) still contain the contribution of numerous unmodelled sources. To understand how this affects the solver (which assumes that the residuals consist of pure noise) we simulate some data and plot a histogram of the real and imaginary parts in Figure \ref{hist}.
If we consider a field consisting of a single 1 Jy source at its centre, the visibilities for this sky have no phase component, and all the visibilities are equal to 1 in this case. The histogram for these visibilities will have two peaks, one at 1 Jy for the real part of the visibilities and the other at 0 Jy for the imaginary component of the visibilities. If we move the source to an offset position from the field centre, the visibilities now have a phase which depends on the offset. Figure \ref{hist1} is the histogram of simulated visibilities for a 1 Jy source at an offset position from the centre. The distribution in this case also has two peaks, one at 1 Jy and one at -1 Jy. In between these peaks, the distribution is almost uniform. 

Figure \ref{hist2} shows the histogram of the visibilities for a field consisting of 100 sources having uniform positions and fluxes drawn from a power law distribution (Pareto distribution with $\alpha=1$) with peak flux set to 1 Jy. Clearly, the distribution of the visibilities approaches a Gaussian. This implies that, given a sufficiently large number of unmodelled point sources with random positions and fluxes (as is usually the case for the fainter sources which do not end up in the model), the distribution of residuals remains Gaussian. Unmodelled point sources therefore tend to simply increase the variance of the residuals above that of the thermal noise contribution. This suggests that unmodelled sources with a flux level below a certain noise-dependent threshold will have almost no effect on the gain solutions. 

Since a typical selfcal procedure begins by constructing a sky model from a 1GC-calibrated image, down to a certain flux threshold, the unmodelled source fraction will tend to consist of multiple faint sources, and therefore will follow Fig.~\ref{hist2}. The properties of the modelled source fraction will, on the other hand, strongly depend on the spatial distribution of flux across the field. Here, we can identify two contrasting regimes. In a field dominated by a bright source, most of the modelled flux will be concentrated in that source, and the distribution of model visibilities will look like Fig.~\ref{hist1}. We'll call this the \emph{concentrated model} regime. In a field with no bright sources, model flux will be spread between multiple fainter sources (we'll call this a \emph{dispersed model} regime), and the visibility distribution will resemble that of Fig.~\ref{hist2}.

We can define the effective SNR of the sky model in terms of the visibilities, as

\begin{eqnarray}
\text{SNR} \, = \, 10\,\text{log}\left(\frac{\langle\V \cdot \V^*\rangle_{\nu,t,pq}}{\langle\N'\cdot \N^{'*}\rangle_{\nu,t,pq}}\right), \label{eq:snr}
\end{eqnarray}

\noindent where $\langle \rangle_{\nu,t,pq}$ denotes averaging over frequency, time and baseline, $\V$ represents the modelled visibilities, and $\N'$ is the effective noise, i.e. the sum of the unmodelled visibilities and the noise. Clearly, SNR is a function of both the total model flux, and the model concentration. For a maximally concentrated model consisting of a single source, $\V \cdot \V^*$ will be equal to the source flux squared. For a disperse model, $\langle \V \cdot \V^* \rangle$ will contain contributions from many interfering fringes. A dispersed model with the same total flux will therefore have much lower SNR. We can then ask whether model concentration, as well as  SNR, affects the degree of source suppression.

Conventional intuition for the workings of selfcal is honed in the ``classic regime'' of high-SNR, concentrated models, typically associated with targeted observations of individual sources. With the advent of blind large-area surveys, we are seeing more and more fields lacking a dominant source: these need to be calibrated in a low-SNR, dispersed model regime. Finally, direction-dependent calibration deals with concentrated models almost by definition, but these can be quite low SNR. The next section shows marked differences in flux suppression across these regimes.

\subsection{SNR, model concentration, and flux suppression}
\label{ssec:AS_SNR}

In this section, we investigate how flux suppression of direction-independent calibration behaves with varying effective SNR and model concentration. We determine under which circumstances we can expect the robust solver to deliver an improvement over the traditional solver. To do this, we simulate a series of observations containing point sources and thermal noise. The point sources are split between a fainter ``unmodelled fraction'' (i.e. assumed unknown for the purposes of calibration), and a brighter known fraction (i.e. included in the calibration sky model). We calibrate the mock observations using the calibration sky model, and then measure flux suppression at the position of the unmodelled sources.
More specifically:

\begin{enumerate}
    \item For the unmodelled source fraction, we generate a sky model containing 100 random point sources as before, and then rescale the fluxes so that their total flux comes to 1 Jy. We call this the faint sky.
    \item For the modelled fraction, we generate a variety of calibration sky models corresponding to different model concentrations and effective SNR levels:
    \begin{description}
    \item[(ii-a)] We fix the total flux in the calibration modelled fraction at 1 Jy, and vary the number of sources from 1 to 50. This corresponds to diluting the model and decreasing SNR simultaneously.
    \item[(ii-b) Concentrated model, varying SNR:] we use a modelled fraction of one source, and scale its flux to achieve different SNR levels. 
    \item[(ii-c) Dispersed model, varying SNR:] we use a modelled fraction of 50 sources, and scale their fluxes to achieve different SNR levels.
    \item[(ii-d) Fixed SNR, varying concentration:] we generate models of 1, 10, 20, 30, 40 and 50 sources. We scale the fluxes of each model to achieve an effective SNR of 10 dB.
    \end{description}
    
    \item We combine the faint sky and the calibration sky model for each experiment, and simulate visibilities corresponding to the combined sky model using the MeerKAT \citep{jonas2018meerkat} array layout. We simulate a single-channel observation at 1 GHz,  with a bandwidth of 1 MHz, a total synthesis time of 2 hours, and an integration time of 10 seconds. 
    \item We add Gaussian noise with an rms of 10 mJy to the simulated visibilities. This value is approximately 3 times the expected rms using MeerkAT system equivalent flux density (SEFD) at frequencies around 1 GHz. This value will be used in all simulations unless stated otherwise.
    \item We perform DI calibration on the data with all three solvers using only the calibration sky model to compute the model visibilities. Since no gains are applied to the visibilities during the simulation, we expect a perfect calibration to return unity gain solutions.
    \item We compute the residuals (by applying the gain solutions to the model visibilities and subtracting them from the data) and image these to get a residual image.
    \item We deconvolve the resulting images using WSCLEAN \citep{offringa2014wsclean} in single scale mode with natural weighting to try and recover the faint source distribution. %{\bf LB - what did you use for the deconvolution? We might want to make sure that the sources fall exactly on pixel centres or use the direct inversion method I showed you.}
\end{enumerate}
We are now in a position to to study the degree of flux suppression of the unmodelled sources. Since our simulations consist of point sources only, the recovered fluxes are estimated by simply measuring the pixel values at the position of the sources in the respective deconvolved images. 

To quantify how a reduction in SNR affects source suppression, we have to create a statistic to measure it with. For this purpose we use the average suppression (AS), which is defined as
 \begin{eqnarray}
\text{AS} \, = \, \frac{1}{N_s}\sum_{i}^{N_s}{\frac{|I_i - \hat{I}_i|}{I_i}}, \label{average_supp}
\end{eqnarray}
where $N_s$ is the number of sources and $I_i$ and $\hat{I}_i$ are the true and recovered flux of the $i^{th}$ source respectively. 

\begin{figure*}
\begin{subfigure}{.48\textwidth}
  \includegraphics[width = 0.98\linewidth]{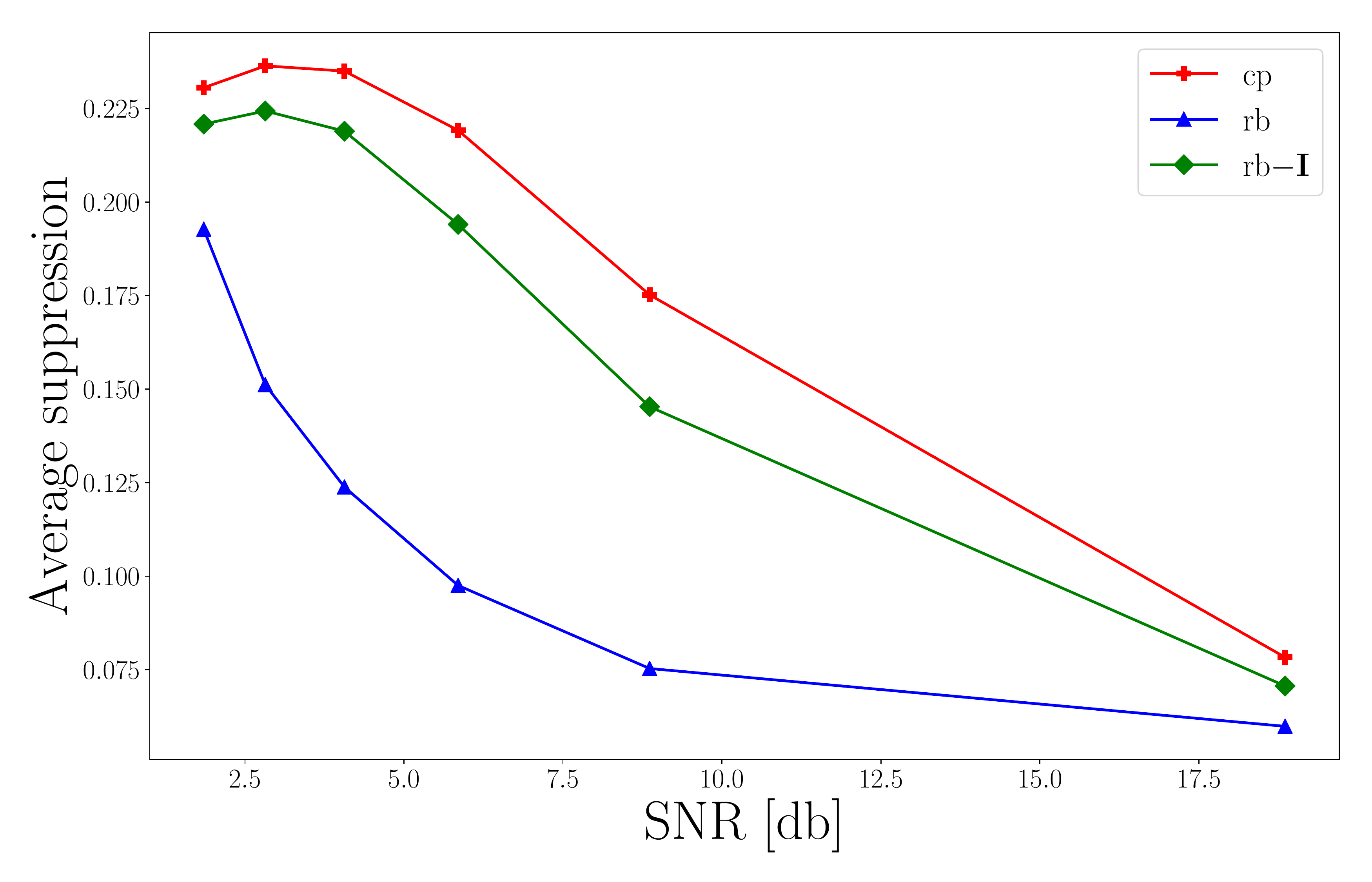}
\caption{Varying SNR and number of model sources} \label{di_supp_snr-a}
 \end{subfigure}
\begin{subfigure}{.48\textwidth}
  \includegraphics[width = 0.98\linewidth]{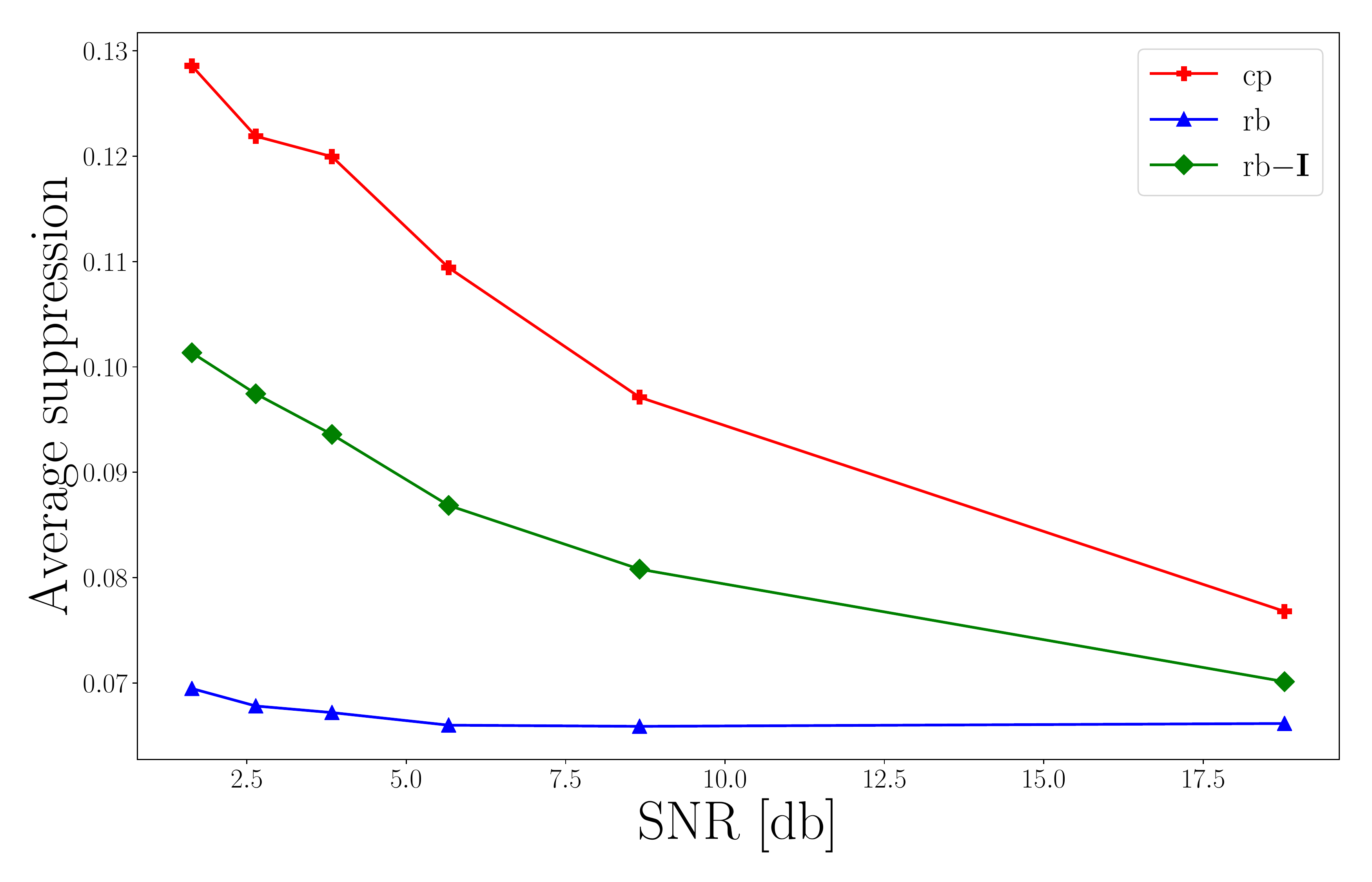}
\caption{1 source model}
\label{di_supp_snr-b}
 \end{subfigure}
\begin{subfigure}{.48\textwidth}
  \includegraphics[width = 0.98\linewidth]{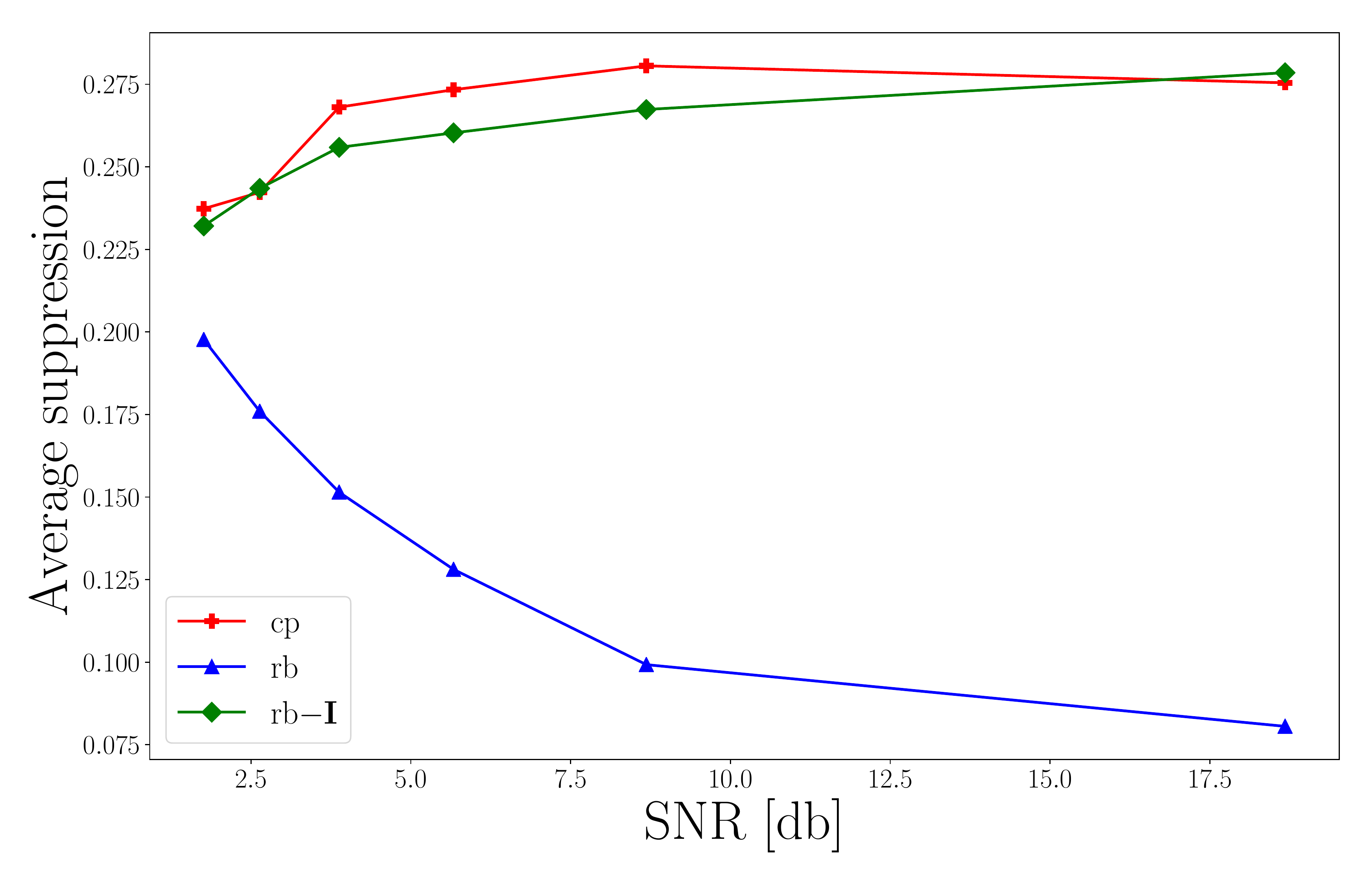}
\caption{50 sources model}
\label{di_supp_snr-c}
 \end{subfigure}
 \begin{subfigure}{.48\textwidth}
  \includegraphics[width = 0.98\linewidth]{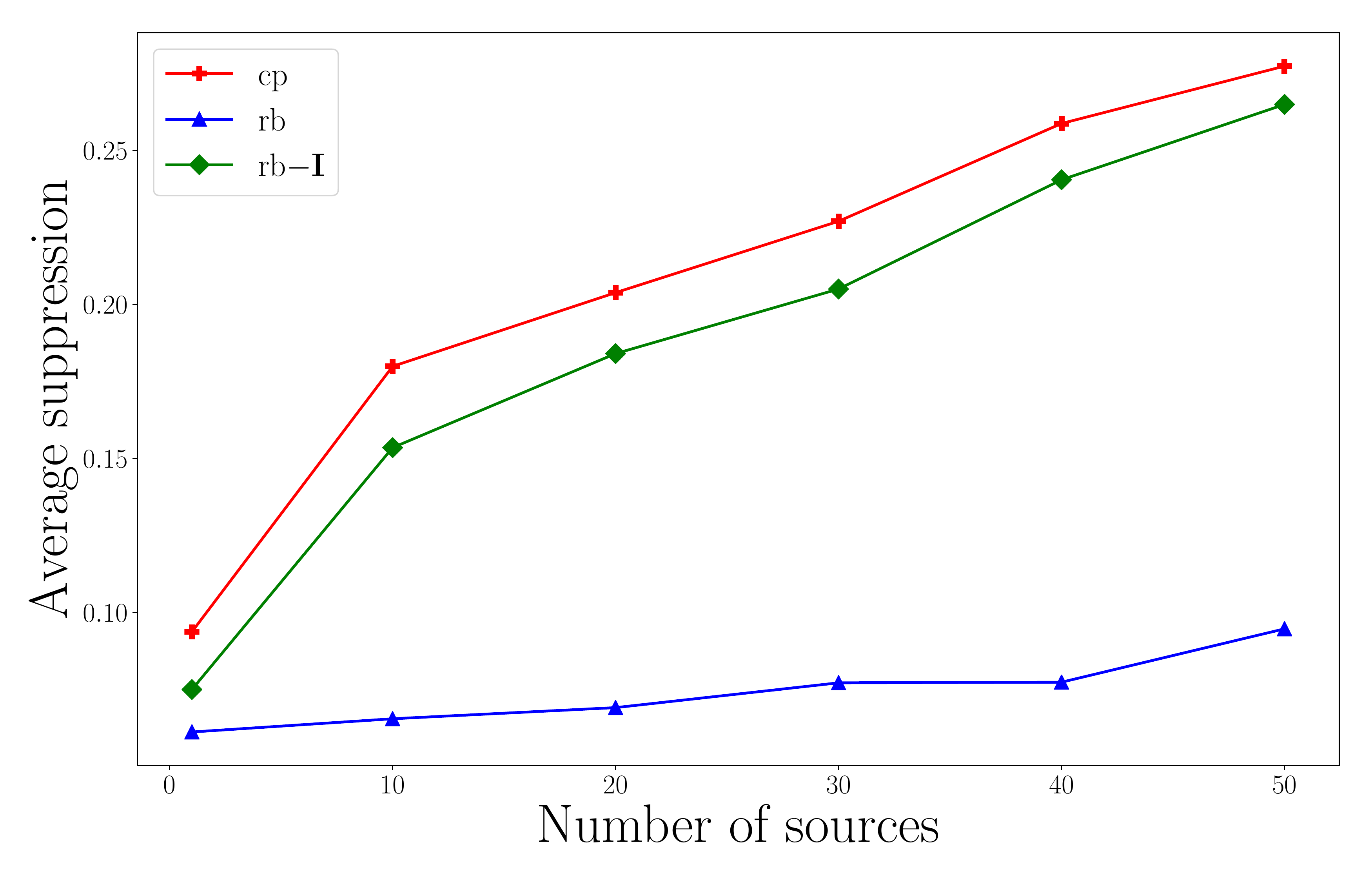}
\caption{Fixed SNR and varying number of model sources}  \label{di_supp_snr-d}
\end{subfigure}

\caption{
\label{di_supp_snr}
The average flux suppression for all the sources in different simulations against the SNR of the data or the number of sources in the model. The red curve is for the complex solver, the blue curve is for the robust solver, and the green curve is the \robustI\ solver.}
\end{figure*}

Figure \ref{di_supp_snr} shows the AS as a function of SNR and model diluteness, following scenarios (a) -- (d) outlined above. These plots reveal a number of very interesting trends:

\begin{itemize}

\item The robust solver (blue) curve always outperforms (in the sense of reducing flux suppression) both the \robustI\ (green curve) and standard solvers (red curve), in some regimes by very a significant margin. The \robustI\ solver outperforms the standard solver in almost all cases, but this improvement is not always significant.

\item For a concentrated model (Fig. \ref{di_supp_snr-b}), flux suppression increases with decreasing SNR. At high SNR (the ``classical regime'' of selfcal), the performance of all solvers tends to converge (Figs. \ref{di_supp_snr-a} and \ref{di_supp_snr-b}, right end of the plot), to a value of slightly below 7\%.\footnote{Previous studies \citep{grobler2014calibration,nunhokee2015link} have shown that flux suppression is highly dependent on array layout and other factors, so the particular value of 7\% is only significant to this series of simulations.}

\item Flux suppression increases significantly (to over 25\%!) with model dispersion (Fig. \ref{di_supp_snr-d}), at least with the standard and \robustI\ solvers. 

\item For a highly dispersed model (Fig. \ref{di_supp_snr-c}), flux suppression with the standard and \robustI\ solvers is quite high, and almost independent of SNR. The robust solver offers much better performance in all but the lowest SNR regimes.

\item There is an interesting downturn in flux suppression at low SNR in Figs.~\ref{di_supp_snr-a}, \ref{di_supp_snr-c} (left end of the plots). We can only speculate as to its ultimate cause. \citet{grobler2014calibration} showed that flux suppression comes about through a combination of ghost sources (see e.g. Eq. 35 therein), and that the intensity of the ghost response has a complex relationship to modelled/unmodelled flux ratios, even in the simplest, two-source case
studied in that work. Perhaps pertinently, Fig.~15 \emph{ibid.} shows a distinct downturn in the ghost response towards low SNR (i.e. higher flux ratios in the figure). We speculate that we are seeing the same mechanism at work here. Furthermore, from continuity considerations it is obvious that there \emph{must} be a downturn in flux suppression at very low SNR -- after all, an empty calibration model cannot suppress flux at all. Since calibration in such a low SNR scenario is pointless, we won't pursue this puzzle further here.

\end{itemize}

The crucial conclusion of this section is that, in principle, the robust solvers outperform the traditional complex solver in all the cases we have considered (at least as far as flux suppression is concerned). The actual degree of improvement is highly dependent on model concentration and SNR. In the extreme regimes, the performance of the solvers appears to converge, so a robust solver may not be worth the extra computational cost. However, as we illustrate in the next section, robust calibration is particularly important for DD calibration, where we are unlikely to operate in a high-SNR regime.

%\OMS{As we illustrate in the next section, this is particularly important for DD calibration, where we are unlikely to operate in a high-SNR regime.}

%\OMScom{In the figure, the solvers converge at very low SNR. So it's a bit confusing that in 4.3 we talk of "high SNR" and "low SNR". It would be good if we could relate the two regimes in 4.3 to an actual SNR level in the figure here, to understand why we're seeing the performance we're seeing.}

\subsection{Flux suppression in DD calibration}
\label{ssec:dd_flux_supp}
Section~\ref{ssec:AS_SNR} shows that the robust solver  significantly improves calibration in a low SNR regime and with a concentrated model. 
%\OMScom{We need to be careful with these statements, and dance around the issue of the solvers working the same when SNR is really low.} 
This is expected to to be the case for DD calibration, since direction-dependent model components tend to be both concentrated, and low in SNR. Consequently, in this section, we extend the simulations in Section~\ref{ssec:AS_SNR} to DD calibration.

\subsubsection{Simulation setup}
\label{ssec:simulated}
We perform two simulations with a similar setup to \cite{kazemi2013robust}, illustrating two characteristic regimes of the solvers. The difference between the simulations is the flux level of the sources relative to the thermal noise and the flux level of the unmodelled sources. Henceforth we refer to them as \emph{high-SNR} and \emph{low-SNR}.

The data were simulated using the same setup as before i.e. MeerKAT array configuration with a single frequency channel at 1 GHz with 1 MHz bandwidth, an integration time of 10 seconds and total synthesis time of 2 hours. For the high-SNR simulation, the noise added to the visibilities has an rms of $10$ mJy which results in an image noise rms of $6\, \mu\text{Jy/beam}$ using natural weighting. For the low-SNR simulation, we add noise with an rms of $0.1$ mJy ($0.06\, \mu\text{Jy/beam}$ image noise rms). 

For both simulations, we generate sky models containing 100 sources with the positions and fluxes generated as before. In the high-SNR simulation, we scale the fluxes of the sources such that the brightest source has a flux of 20 Jy, while for the low-SNR simulation, we scale the fluxes so that the brightest source has a flux of $0.05$ Jy. We choose a peak flux of 20 Jy for the high-SNR regime in order to replicate one of the setups in \citet{kazemi2013robust} where the modelled sources are very bright (i.e.> 5 Jy) and the unmodelled sources are also relatively bright (reaching values even up 3 or 4 Jy). This simulation is similar to the high-SNR end of Fig.~\ref{di_supp_snr-b}. In the low-SNR simulation, we seek to investigate a different regime where fluxes of model sources are very low and comparable to the faint sky. Here we expect a scenario similar to the low SNR part of Fig.~\ref{di_supp_snr-b}.
%\OMScom{We need to motivate the choice of values, or at least relate these to the SNR values in Fig. 2.}

We assume that the 10 brightest sources are included in the calibration model, and the remaining 90 are unmodelled. We corrupt the 10 brightest sources with DD gains (technically, such gains will affect all sources and not just the brightest ones, but for reasons of computational economy, we restrict DD gains to the modelled sources) and add Gaussian noise to the corrupted visibilities. We apply smoothly varying DD gains generated from a circularly symmetric Gaussian process\footnote{A Gaussian Process (GP) (see \cite{rasmussen2006gaussian}) is a Gaussian distribution over functions. The complex gains' realisations have a constant mean function $1 + 0j$, thus constraining the gains to mean amplitudes $1$ and phases $0$. The squared exponential covariance function, $k$, is used to control the smoothness and variability of the gain. It is defined for two adjacent points $x$ and $x'$ as 
\begin{eqnarray}
k(x,x') \, = \, \sigma_{f}^2 \exp \frac{{(x-x')^2}}{2l^2}, \label{SE_eqn}
\end{eqnarray} 
where the hyperparameters of the covariance function, $\sigma_{f}$ and $l$ stand for the standard deviation of the GP and the length scale (i.e. the input length for the GP to vary considerably) respectively. We set the value of $l$ to 300 (i.e. 30 units of the integration time), and $\sigma_{f}$ to $0.1$.}

\subsubsection{Results}
\label{ssec:dd}

We perform DD-calibration on the corrupted visibilities, with only the 10 brightest sources modelled, using a solution interval of $150 \,\text{s.}$ Calibration is performed using the complex, robust and \robustI\ solvers. As in the previous simulation, after calibration, we produce residual images to study the suppression in the fluxes of the unmodelled sources. 

\begin{figure*}
\begin{subfigure}{.48\textwidth}
  \includegraphics[width = 0.99\linewidth]{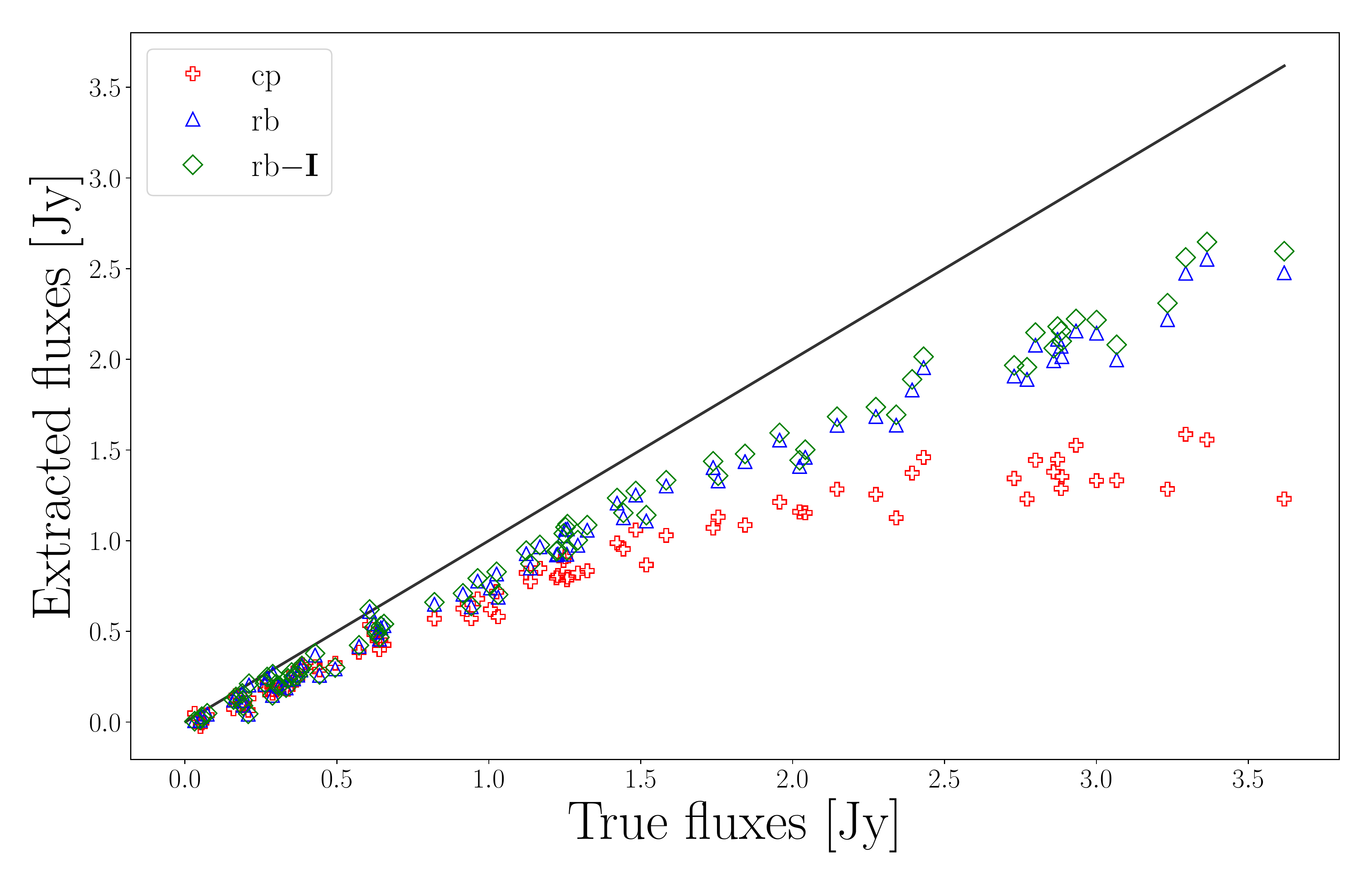} 
\caption{}\label{I20pix_ratio_tint_15}
 \end{subfigure}
\begin{subfigure}{.48\textwidth}
  \includegraphics[width = 0.95\linewidth]{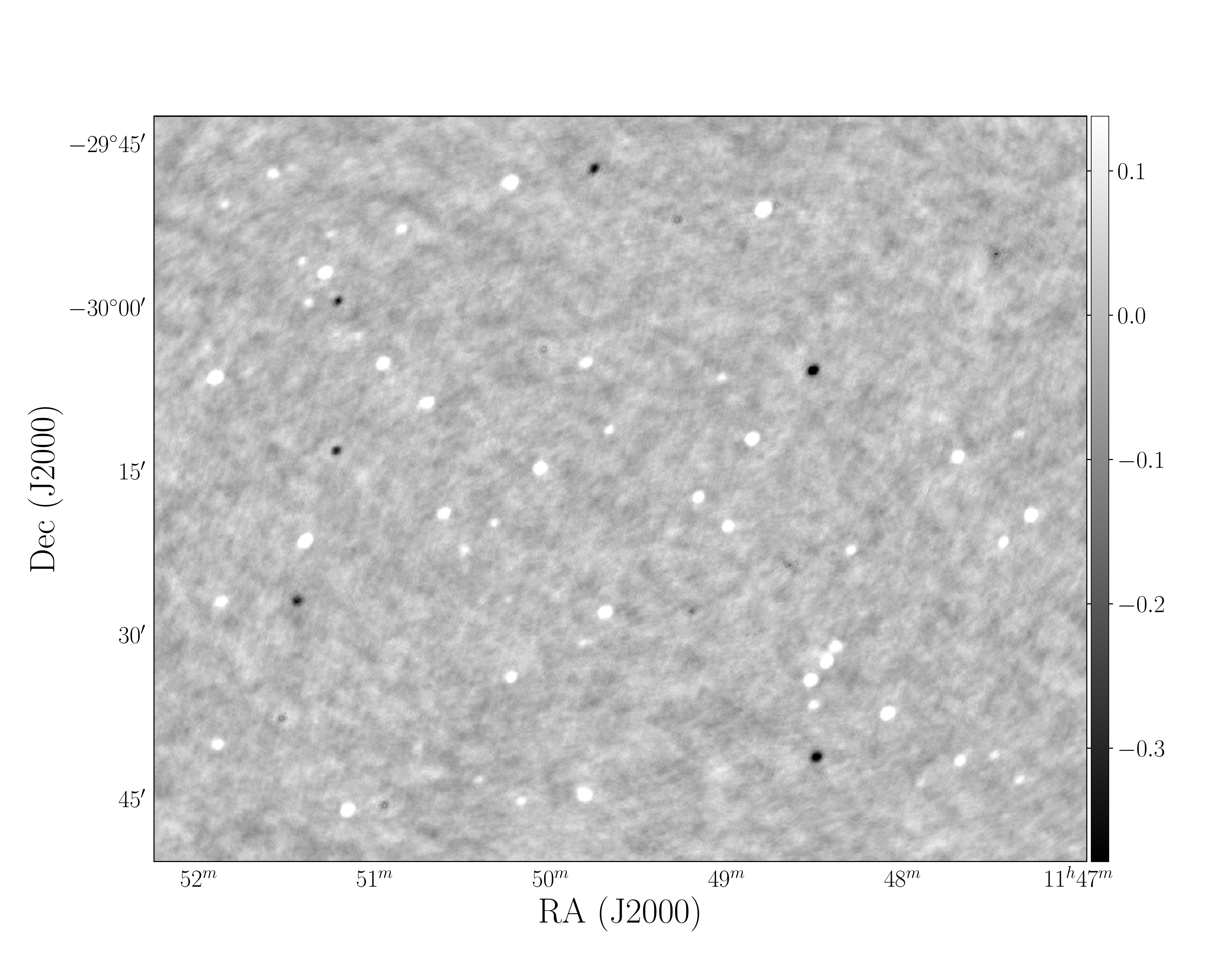}
 \caption{}\label{I20Jy_diff_tint_15}
 \end{subfigure}
\caption{(a): Recovered flux against input flux for the high-SNR simulation. (b): Difference map between the corrected residuals of the \robustI\ solver and the complex solver.}
\label{I20Jy_tint_15_flux}
\end{figure*}

\begin{figure*}
\begin{subfigure}{.48\textwidth}
  \includegraphics[width = 0.99\linewidth]{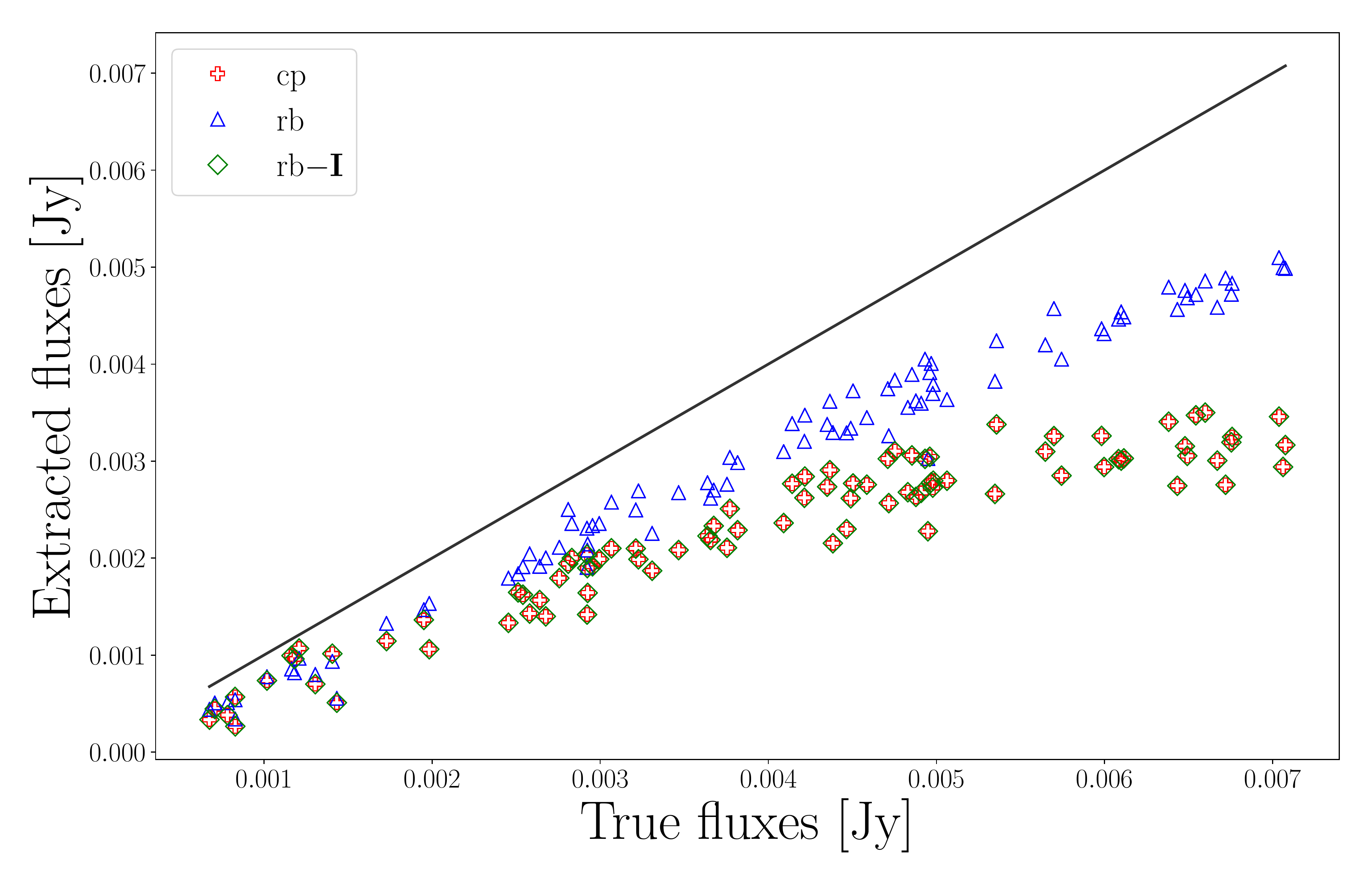} 
\caption{}\label{I1Jypix_ratio_tint_15}
 \end{subfigure}
\begin{subfigure}{.48\textwidth}
  \includegraphics[width = 0.95\linewidth]{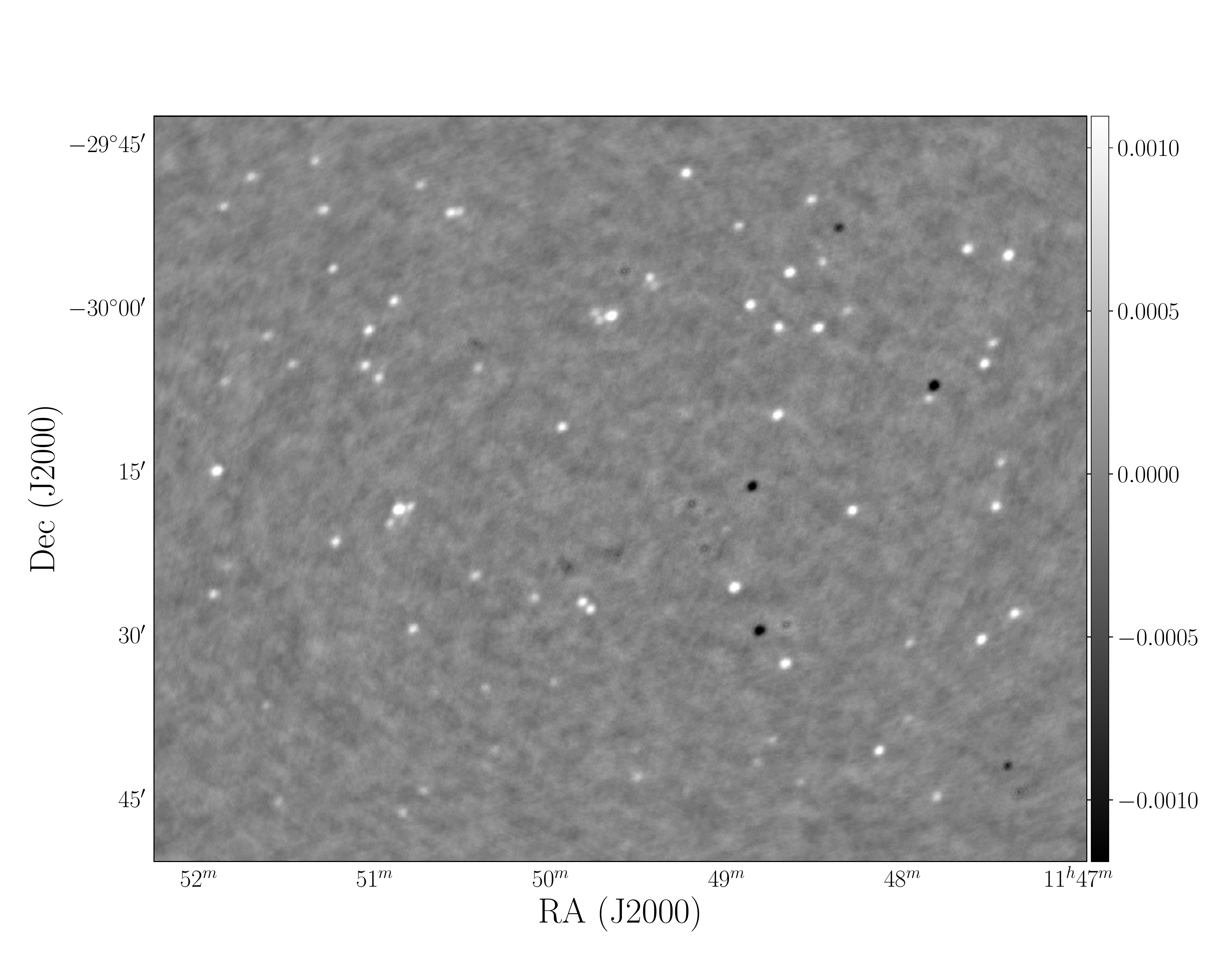}
\caption{} \label{I1Jy_diff_tint_15}
 \end{subfigure}
\caption{(a): Recovered flux against input flux for the low-SNR simulation. (b): Difference map between the corrected residuals of the robust solver and the complex solver.}
\label{I1Jy_tint_15_flux}
\end{figure*}

We show the results of the high-SNR simulations in Figure~\ref{I20Jy_tint_15_flux}. Figure \ref{I20pix_ratio_tint_15} shows the recovered flux against the input flux for the different algorithms. We observe that both the robust solvers out-perform the complex solver with the \robustI\ solver producing marginally better results than the robust solver. We show the difference map, i.e. the image recovered by the \robustI\ solver, minus the image recovered by the complex solver in Figure \ref{I20Jy_diff_tint_15}. The difference image has numerous bright, positive peaks corresponding to the additional flux recovered by the \robustI\ solver. The low-SNR simulation, by contrast, shows the robust solver produces the best results (see Figure \ref{I1Jy_tint_15_flux}). The difference maps in Figure \ref{I1Jy_diff_tint_15} further emphasise this. We also note the negative peaks which occur at the positions of the \emph{modelled} sources in both Figure~\ref{I20Jy_diff_tint_15} and Figure~\ref{I1Jy_diff_tint_15}. These peaks imply that the complex solver residuals contain \emph{more} flux at the model source positions. Since the modelled sources (with DD-gains applied) have been subtracted from the residual maps, this, in turn, implies that, in the presence of unmodelled sources, the complex solver tends to \emph{underestimate} the modelled sources to a greater extent than the robust solvers.\\
\indent The reason why the \robustI\ solver performs well in the high-SNR simulation compared to the low-SNR simulation is that for the high-SNR simulation, the covariance of the residuals is higher than $\mathbf{I}$. Hence, in this simulation, visibilities are adequately weighted. However, for the low-SNR simulation, the variance is under-estimated (residual covariance is smaller than $\mathbf{I}$) effectively assigning equal weights to all visibilities and hence results in a similar performance as the complex solver. 


\subsubsection{Solution intervals}
\label{ssec:solints}

One of the most critical decisions during calibration is the choice of solution intervals. Solution intervals are generally employed to improve the SNR and to make the system of equations over determined. Ideally, solution intervals are chosen such that they are shorter than the time and frequency scales of the gains' variability, but long enough to provide significant SNR. For differential gains (or DD calibration), longer solution intervals are thus necessary (since the SNR in per-direction models is lower); somewhat fortuitously, physical intuition suggests that in most regimes, the DD component of the gain (e.g. primary beam rotation) should vary slower in frequency and time relative to the DI component (e.g. atmospheric phase). In order to investigate the effects of solution intervals on gain solutions, we repeat the experiments above while varying the solution intervals. 

\begin{figure*}
\begin{subfigure}{.48\textwidth}
  \includegraphics[width = 0.98\linewidth]{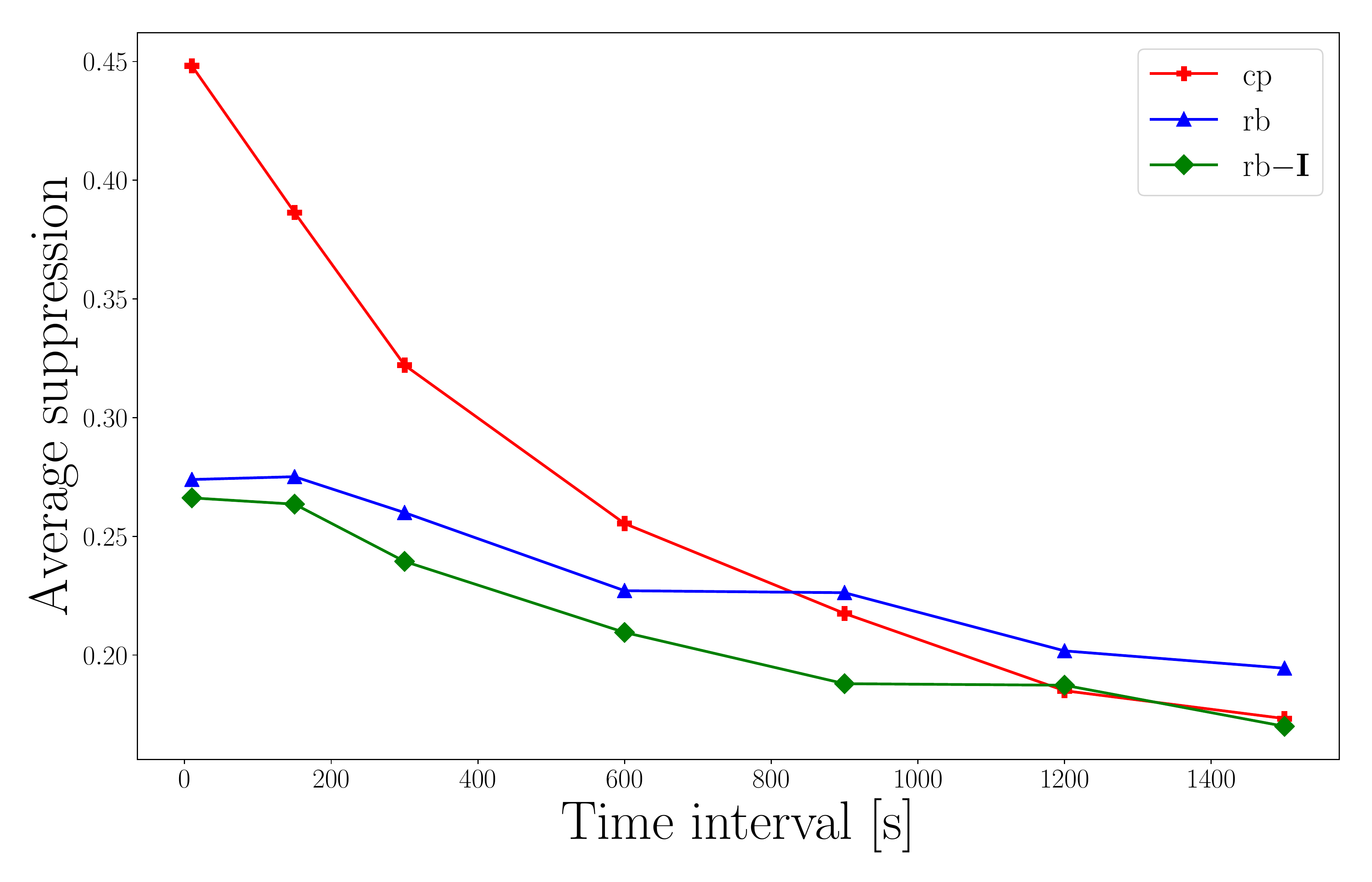}
\caption{} \label{I20Jy_avg_supp_tint}
 \end{subfigure}
\begin{subfigure}{.48\textwidth}
  \includegraphics[width = 0.98\linewidth]{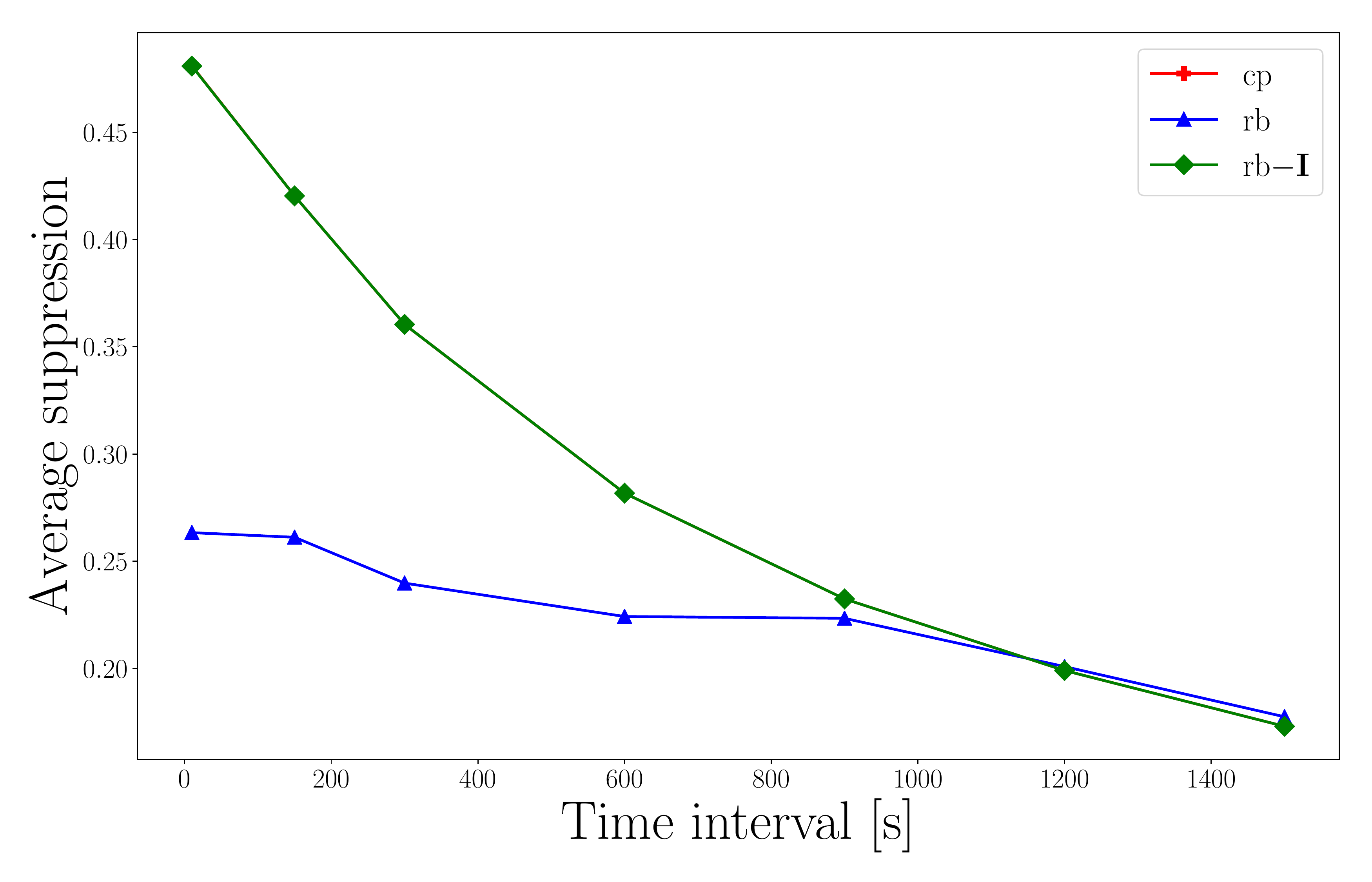}
\caption{} \label{I1Jy_avg_supp_tint}
 \end{subfigure}
\caption{The average flux suppression (AS) across all sources in the simulation. (a): High-SNR regime (b): Low-SNR regime.}
\label{I20_1Jy_avg_supp_tint}
\end{figure*}

The average suppression (AS) as a function of solution interval is shown in Figs.~\ref{I20Jy_avg_supp_tint} and \ref{I1Jy_avg_supp_tint}. The figures show that as the solution interval increases, flux suppression goes down, which is consistent with the results of \citet{nunhokee2015link}. At sufficiently large time intervals, all three solvers eventually reach an asymptotic level of flux suppression.
%in the low-SNR case this level is similar across all three solvers, while in the high-SNR case, the \robustI\ solver performs markedly better than the robust or complex one. \OMScom{An explanation of this is needed.}

This clearly illustrates the benefits of a robust solver (at least in the sense of lower flux suppression) only kick in in specific regimes. In particular, in the low-SNR regime, if the gains are sufficiently stable for long solution intervals to give acceptable results, the (computationally cheaper) complex solver produces almost equivalent results to the [more expensive] robust solvers. With shorter solution intervals, the robust solvers tend to produce markedly lower flux suppression. We note that gain (in)stability is not the only reason to choose shorter solution intervals: there may also be purely operational reasons. In particular, the amount of data produced by new arrays such as MeerKAT (and the future SKA will push this up by orders of magnitude) drives a requirement for data parallelism, while at the same time increasing the memory footprint of existing algorithms. This implies that the data needs to be processed in smaller chunks, thus constraining the size of a practical solution interval for this class of algorithms, and potentially opening a precious niche for robust solvers.\footnote{\color{black}For completeness, we should note other approaches to the small-chunk problem, such as consensus optimisation \citep{10.1093/mnras/stv596}, filtering \citep{tasse2014nonlinear} and recently stochastic LBFGS \citep{yatawatta2019stochastic}.}

\section{Robust solvers and RFI mitigation}
\label{sec:sim_rfi_data}
The robust solver works by iteratively recomputing weights based on how far our modelled visibilities are from the observed visibilities. During calibration, the robust solver will tend to suppress the effect of remaining outliers in data such as those caused by low-level RFI, which is particularly difficult to remove using conventional data flagging. This is conceptually similar to the approach of \citet{bonnassieux2018variance}, where uncertainties from calibration solutions are used as weights during imaging to reduce the effects of outliers. Note that here, however, the weights from the robust solver cannot be used for imaging as these will tend to suppress the unmodelled sources. We demonstrate this behaviour in a simulation, and then on real observational data.

The dataset in question is a 1.2 hours 2013 VLA
observation of the VIDEO deep field (J2000, RA=$02h11m21.09s$, Dec=$-04d11m13.5s$). VIDEO was deliberately chosen as a field relatively free from bright sources (so as to minimise the level of deconvolution and DDE-related artefacts), with the brightest object in the field being only $\approx\,0.02$ Jy. This particular observation covers a frequency range of 0.9--2.6 GHz, with 16 spectral windows each having 64 channels. The integration time on average is 9 seconds. It employs 28 VLA antennas, with a maximum baseline of 36.4 km.  For this experiment, we first transform the measurement set to have a single spectral window by combining all spectral windows. We obtained the data after initial flagging and 1GC calibration using the CASA software (see \citet{IanVIDEO} for more details). We then image the 1GC-corrected data, and extract a component-based sky model using the PyBDSF package \citep{mohan2015pybdsf}. This sky model is used as a basis for the simulations in this section.

Before testing our solvers on real data, we first discuss the qualitative effects of unflagged RFI on data processing, and present some simulations to illustrate our predictions. For the sake of simplicity, we restrict the discussion and our simulations to stationary terrestrial RFI sources; we note, however, that other types of RFI (e.g. self-RFI, aircraft and satellite RFI) also manifest themselves as outliers in the data (see \citet{offringa2015} for a few examples).

\subsection{Simulating low-level RFI}
Let's consider a single narrow-band (and, possibly, on/off or time-variable) RFI source. Stationary (terrestrial) RFI sources are fixed with respect to the baselines, and therefore have a nominal fringe rate of zero. Radiation from a stationary RFI source is (as far as the interferometer is concerned, in a given timeslot and frequency channel, and assuming the receiver chain is not saturated by the RFI signal) indistinguishable from a real source at either celestial pole, modulo the primary beam gains, modulo a constant phase offset. Delay tracking in the correlator, being more rapid for longer baselines, consequently imposes a higher fringe rate for such sources on longer baselines, which attenuates the RFI response on longer baselines due to time and bandwidth averaging. 

If we consider only the imaging problem, the net effect of an (unflagged) RFI source is then very similar to that of a bright source at a celestial pole. Images of the target field will be contaminated by structure that is modulated by the PSF sidelobes of a polar source. For low-level RFI, and a field sufficiently far from a pole, these can be ignored, or even lost in the noise. This is especially true in the case of continuum imaging. One can think of it in terms of \emph{RFI occupancy}: a narrow-band, on/off source contributes to relatively few of the visibilities that go into a Fourier transform (i.e. has low occupancy), therefore its effect on the image is diluted.

The effect on calibration can be far more insidious, particularly if short time/frequency intervals are employed. Within a particular short time/frequency interval, an RFI source can happen to have high occupancy, thus significantly biasing the gain solutions for that interval. In the worst case, the gain solutions are biased low, and applying their inverse then ``blows up'' some of the corrected visibilities. Let's consider the following RIME model as an example
\begin{equation}
\V_{pq} = \G_p \mat{C}_{pq} \G_q^H + \bm{\epsilon}_{pq} + \bm{\eta}_{pq},
\label{rime_simple}
\end{equation} 
where $\V_{pq}$ denotes the corrupted visibilities, $\G_p$ represents the gains for antenna $p$, $\mat{C}_{pq}$ is the sky coherency, $\bm{\epsilon}_{pq}$ and $ \bm{\eta}_{pq}$ are the noise and RFI corruptions respectively. The corrected data after calibration, $\V_{pq}^{c}$ is obtained by applying the inverse of the estimated gains, $\hat{\G}$ to the data as follows
\begin{align}
\V_{pq}^{c} &= \hat{\G}_p^{-1} \V_{pq} \hat{\G}_q^{-H}, \label{corr_c1}\\
&= \hat{\G}_p^{-1} \left(\G_p \mat{C}_{pq} \G_q^H + \bm{\epsilon}_{pq} + \bm{\eta}_{pq}  \right)\hat{\G}_q^{-H}.\label{corr_c2}
\end{align}
Consequently, if the gains of antenna $p$, for example, are biased low by RFI, the application of their inverses not only amplifies the RFI but also the noise. Since the noise is present on all baselines containing antenna $p$, the amplified noise now has a high occupancy, and results in strong imaging artefacts. 

We now perform a simulation in order to illustrate these arguments. We replicate the VIDEO observation by using the MeqTrees package \citep{noordam2010meqtrees} to simulate the visibilities corresponding to the sky model for the VIDEO deep field derived above. We then inject a mock low-level RFI source into the data by simulating a 3.5 Jy point source at the South Celestial Pole\footnote{The flux value was deliberately chosen to illustrate the effects above. Note that our simulation does not include primary beam attenuation, so the quoted brightness of the RFI source is unattenuated.}. We simulate the visibilities corresponding to the RFI source separately, again using MeqTrees, with time and bandwidth smearing enabled, which effectively attenuates power on the longer baselines, as would be expected for a real RFI source. In order to replicate the narrow-band and on-off behaviour of RFI, we inject the simulated RFI visibilities into the simulated sky data at a randomly sampled subset of timeslots and frequency channels. We also add thermal noise at a level of $0.16\text{ Jy}$ (which corresponds to the rms estimated from the real data measurement set). We do not add any other effects to the simulation, as the objective of the experiment is to study the impact of the RFI in isolation.

\begin{figure*}
\begin{subfigure}{.31\textwidth}
  \includegraphics[width=6cm,height=4.8cm,keepaspectratio]{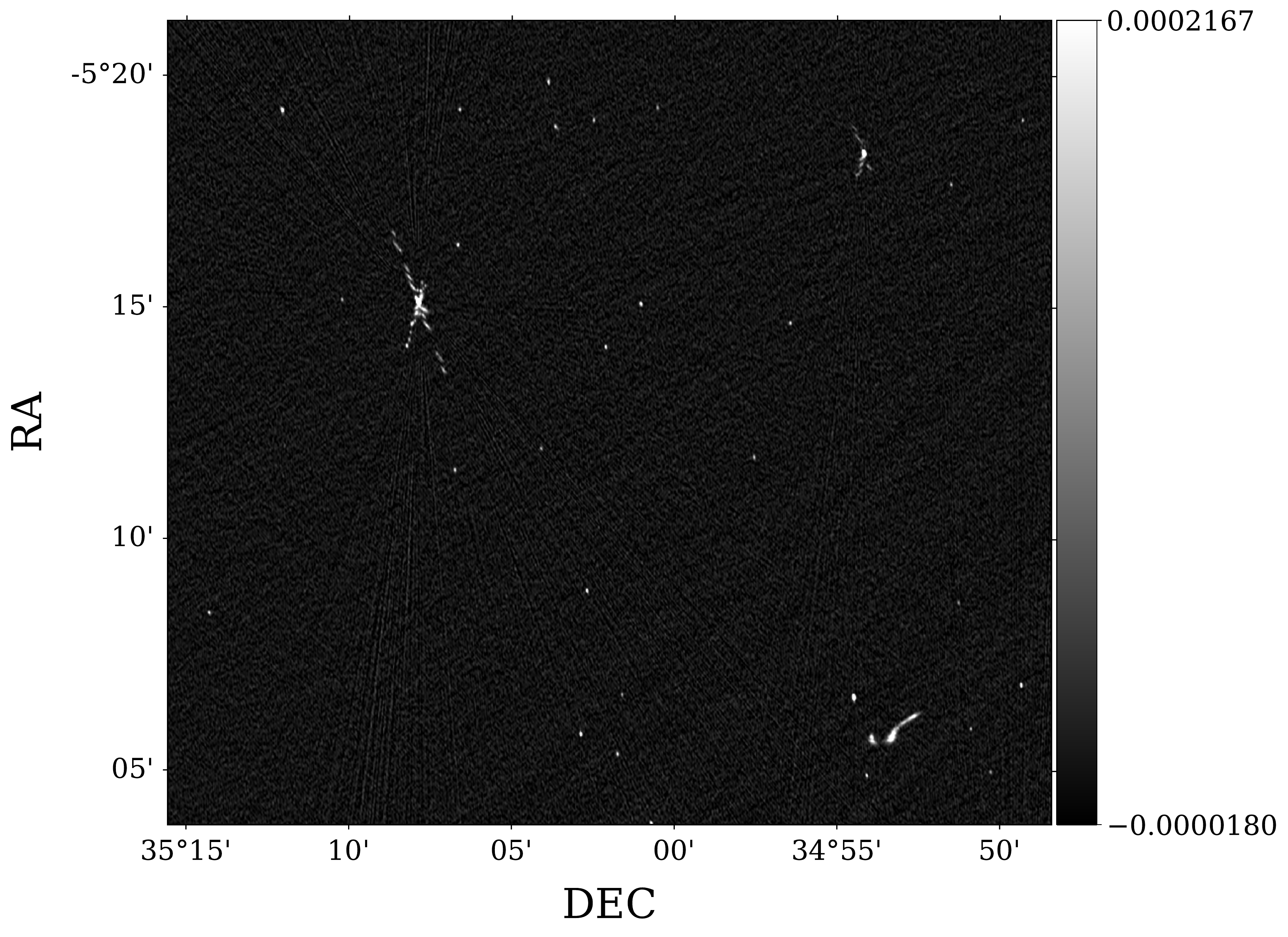} %sim_deep_2_rfi_1GC__image.pdf
\caption{RFI Data}\label{deep_RFI_sim}
 \end{subfigure}
\begin{subfigure}{.31\textwidth}
  \includegraphics[width=6cm,height=4.8cm,keepaspectratio]{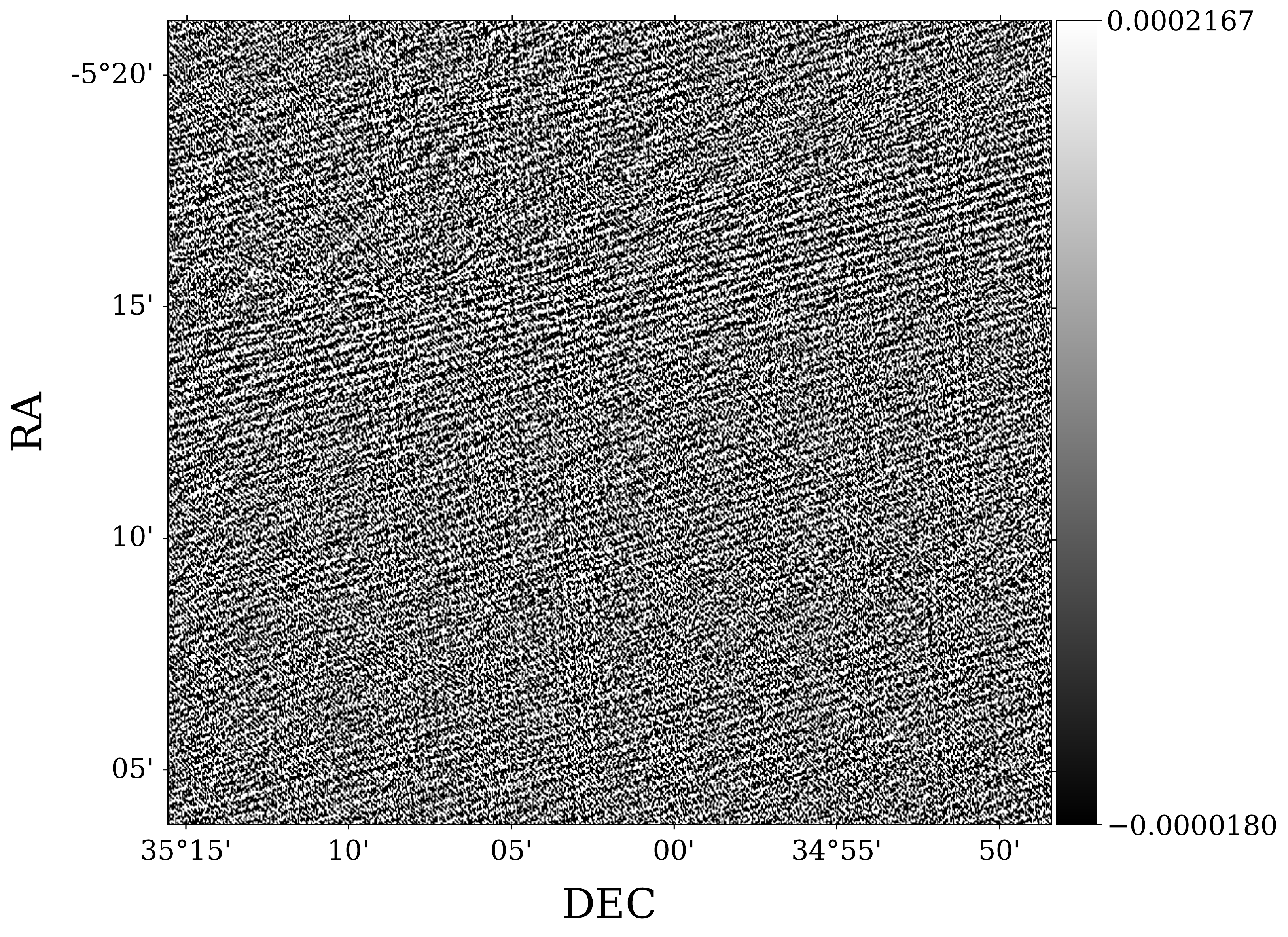} %sim_deep_2_complex_fint_128_tint_1_image.pdf
    \caption{complex solver time-int = 9 secs} \label{sim_complex_tint1}
 \end{subfigure}
%\vspace{-1cm}
\begin{subfigure}{.31\textwidth}
  \includegraphics[width=6cm,height=4.8cm,keepaspectratio]{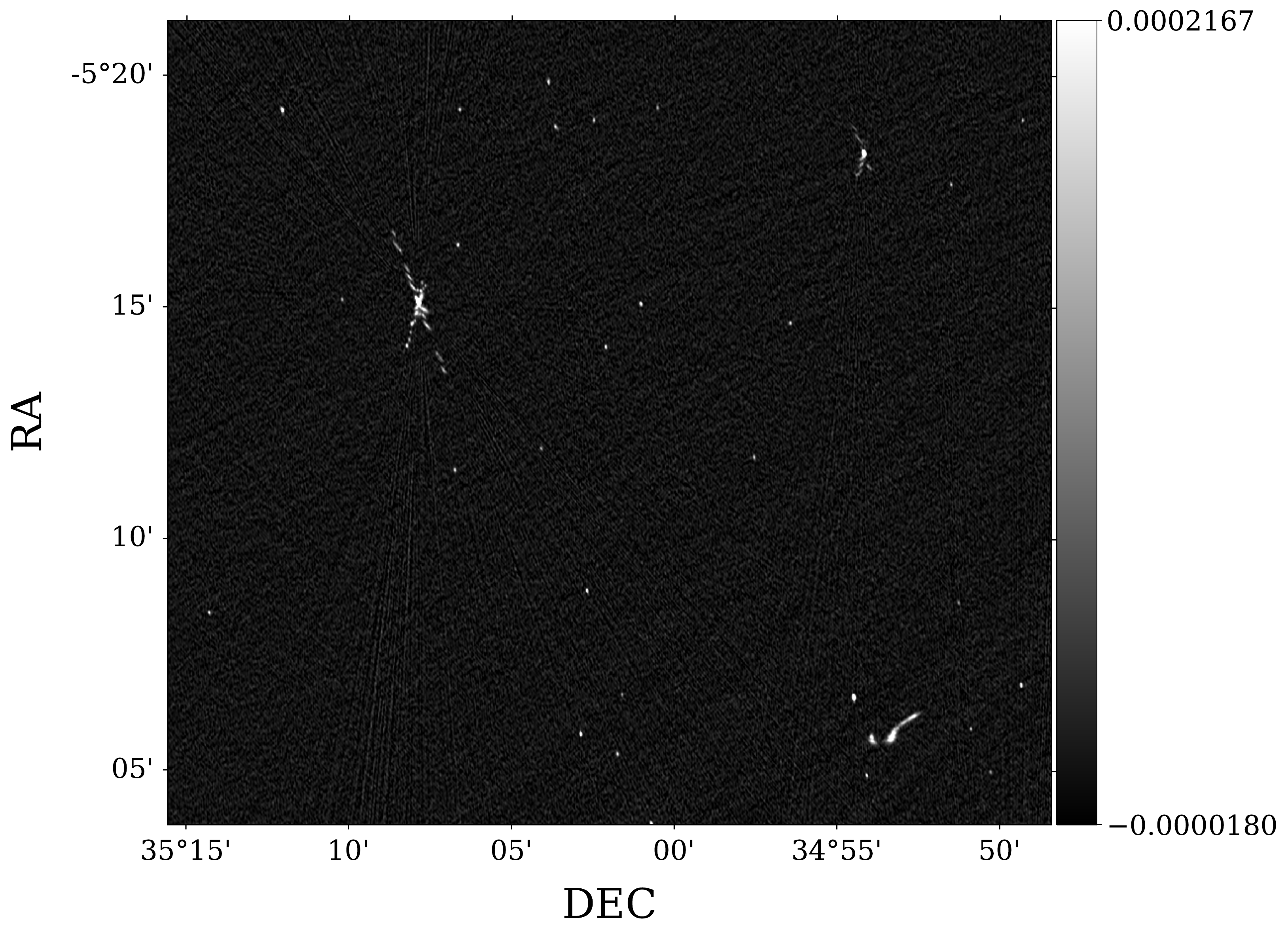} %sim_deep_2_robust_fint_128_tint_1_image.pdf
\caption{robust solver time-int = 9 secs}\label{sim_robust_tint1}
 \end{subfigure}
 \caption{Images of a patch at the centre of the field for simulated RFI-corrupted data, and corrected data after calibration with both solvers. (a) RFI-corrupted data, (b) after calibration using the complex solver with a time interval of 9 secs, (c) after calibration using the robust solver with the same time interval. RFI-induced artefacts are clearly visible in case (b).}
\label{Sim_deep_2_images}
\end{figure*}

Having simulated our mock-RFI-contaminated data, we perform full amplitude and phase DI calibration on the data with both the complex and the robust solver using a time interval of 9 seconds and a frequency interval 128 MHz. Figure \ref{Sim_deep_2_images} shows maps of a field-centre patch of the simulated data, as well as the corrected data obtained after calibration with both solvers. In Figure \ref{deep_RFI_sim}, the RFI source manifests itself as faint linear structure in the background. In Figure \ref{sim_complex_tint1}, some gain solutions from the complex solver have been biased by RFI, as predicted by the discussion above, resulting in significant image degradation. By contrast, with the robust solver (Fig.~\ref{sim_robust_tint1}), no such contamination occurs, as the robust solver effectively excludes the RFI-affected visibilities via its weighting scheme. 
%{\bf LB - The editor might complain about how you are referencing the sub-figures. Can't you reference then directly?}

\subsection{Application to real data}
\label{ssec:real_data}
\begin{figure}
 %\hspace*{-0.85cm}
  \includegraphics[width = 0.5\textwidth]{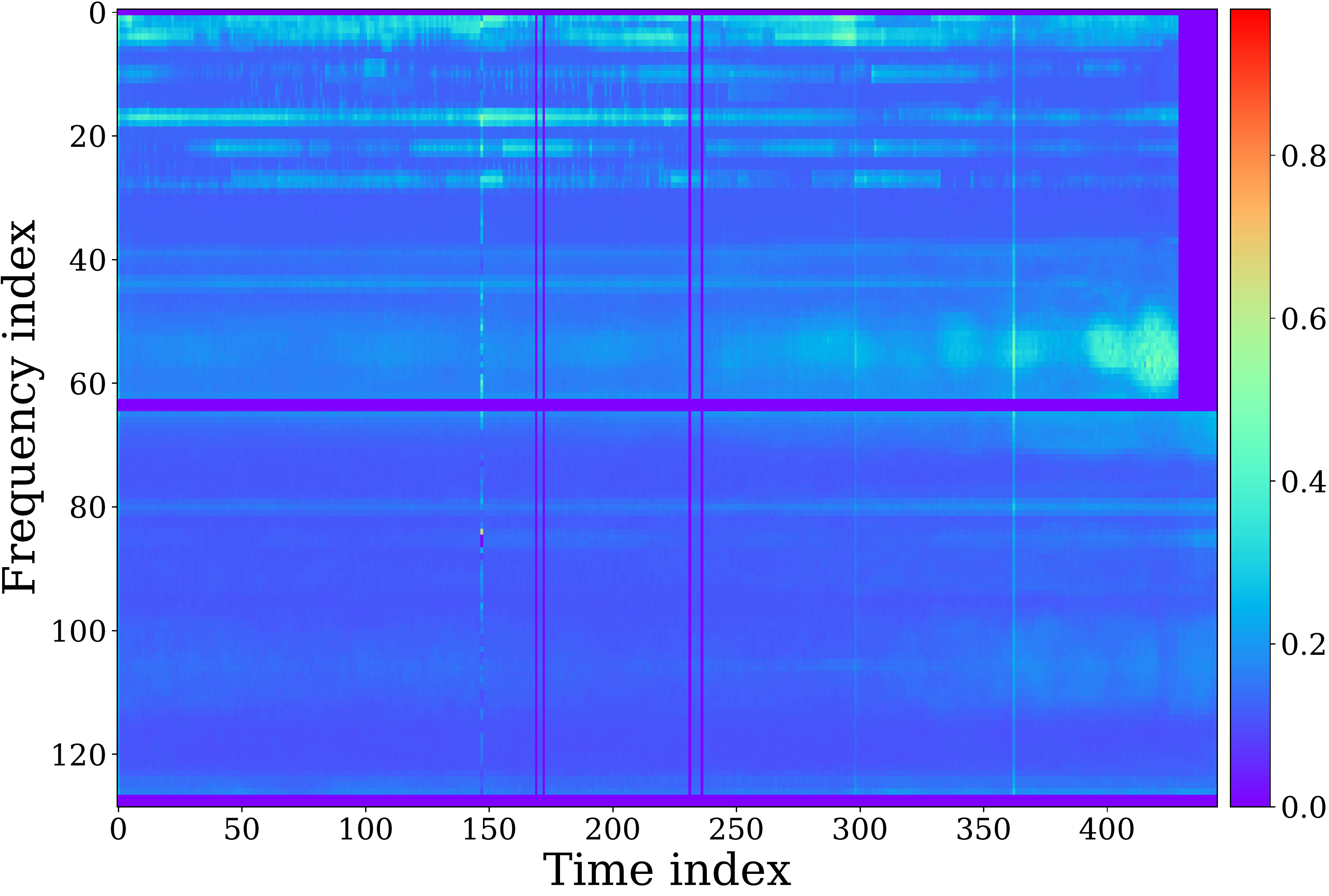}
  \caption{A waterfall plot of the  visibilities for the 
  VIDEO observation before self-calibration, averaged across all baselines and correlations. This is an image of a chunk of data containing 128 frequency channels. The gains plotted in Figure \ref{deep_gains} correspond to the same data chunk. The purples stripes correspond to previously  flagged data, and the bright spots correspond to low-level RFI.}
\label{deep_2_rfi}
\end{figure}

Having demonstrated the success of the robust solver on simulated data we now attempt to calibrate the real VIDEO data set. This data set is an excellent test case because it is a deep field with low SNR and the data is contaminated by low-level RFI which is difficult to remove using conventional flagging. Figure \ref{deep_2_rfi} is a waterfall plot of the average visibilities. The bright spots are the visibilities which are corrupted by the low-level RFI. Since the data was already bandpass calibrated during 1GC, we perform self-calibration with solution intervals of 9 sec and 7.5 min, using both the robust and the complex solver. 

\begin{figure*}
\begin{subfigure}{.48\textwidth}
  \includegraphics[width=8cm,height=6.5cm,keepaspectratio]{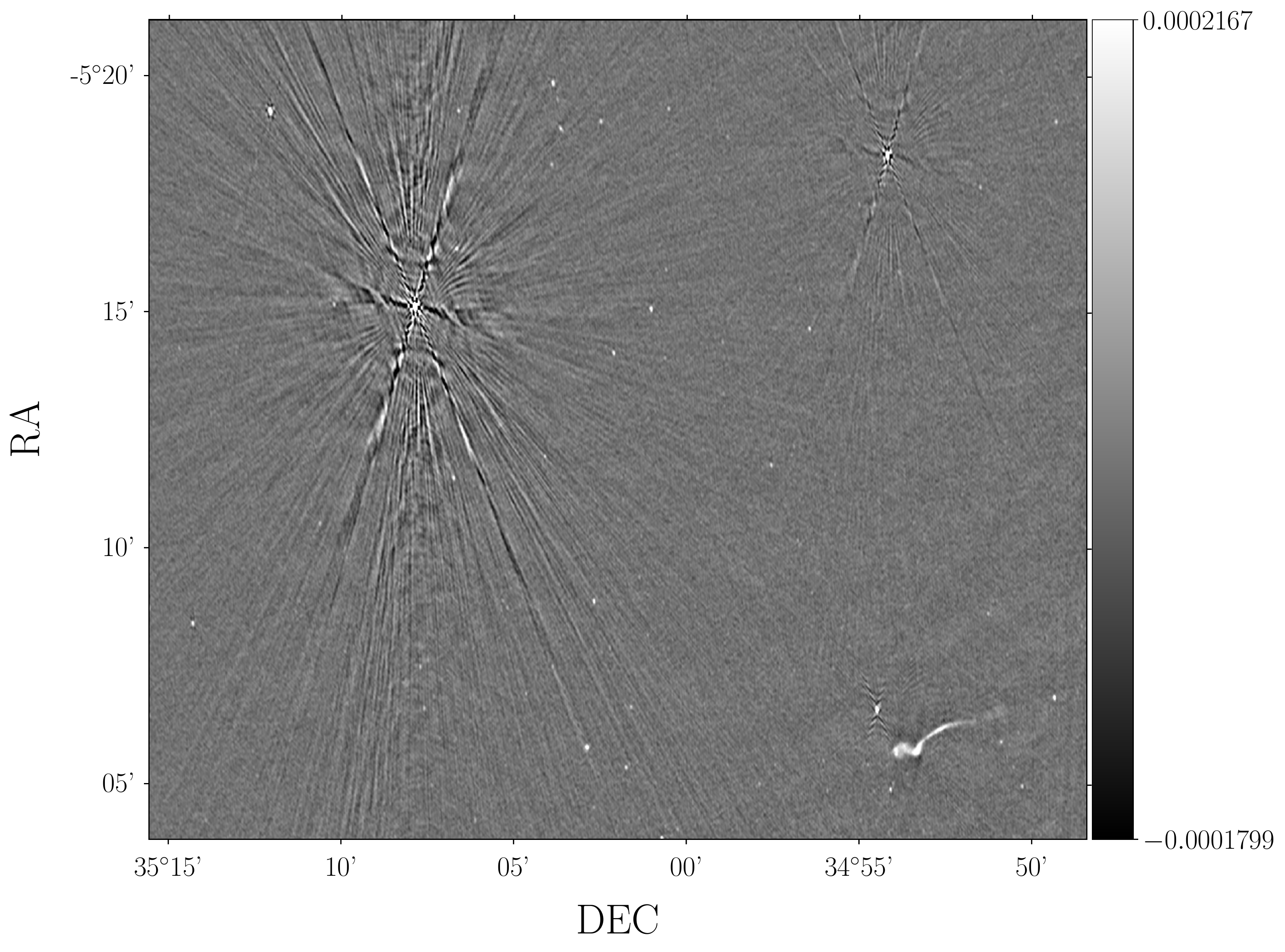} %deep_2_1GC__image.pdf
\caption{Uncalibrated data}\label{deep_befor}
 \end{subfigure}
\begin{subfigure}{.48\textwidth}
  \includegraphics[width=8cm,height=6.5cm,keepaspectratio]{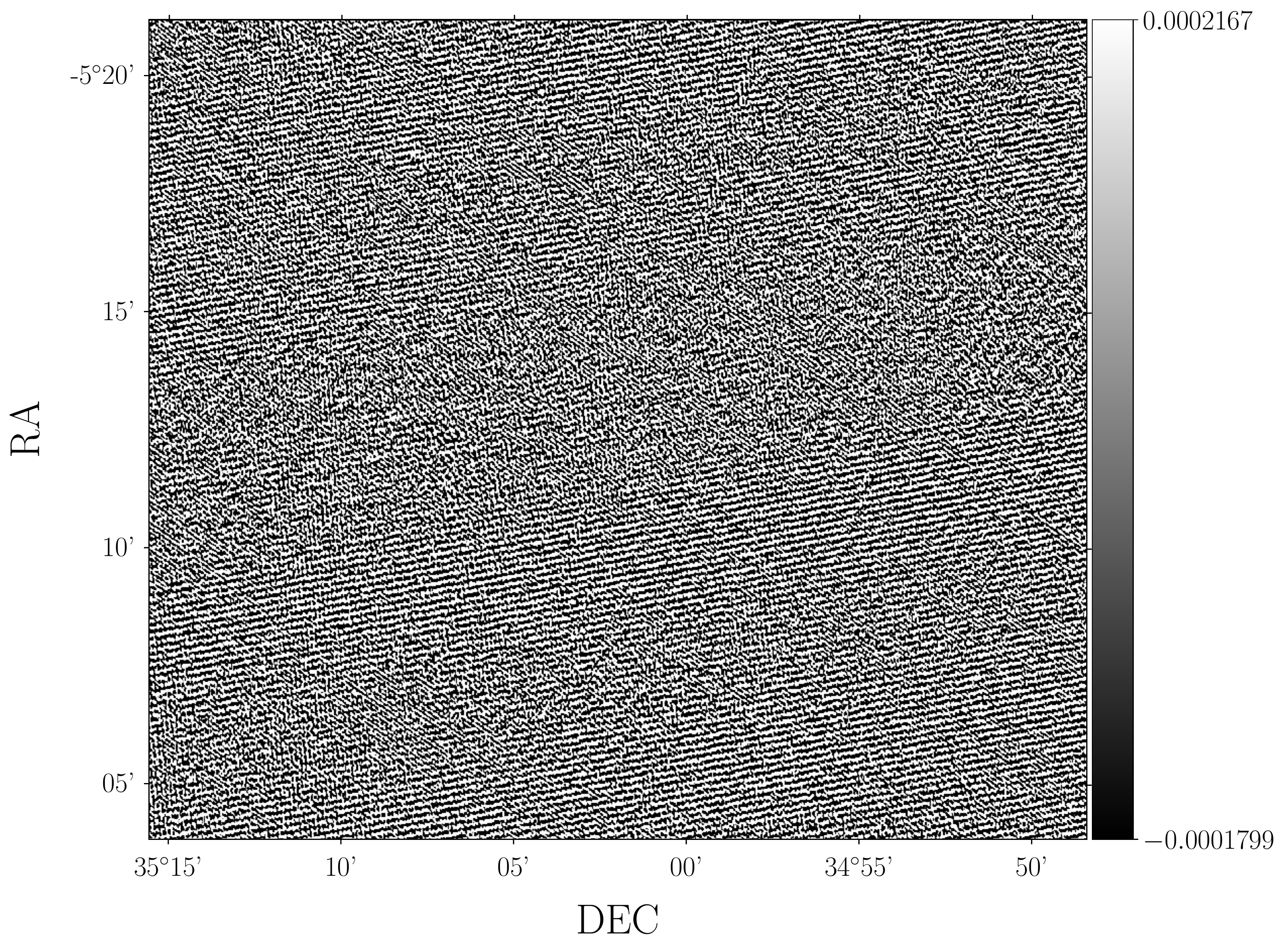} %deep_complex_fint_128_tint_1_image.pdf
    \caption{Complex solver, 9 sec intervals} \label{complex_tint1}
 \end{subfigure}
%\vspace{-1cm}
\begin{subfigure}{.48\textwidth}
  \includegraphics[width=8cm,height=6.5cm,keepaspectratio]{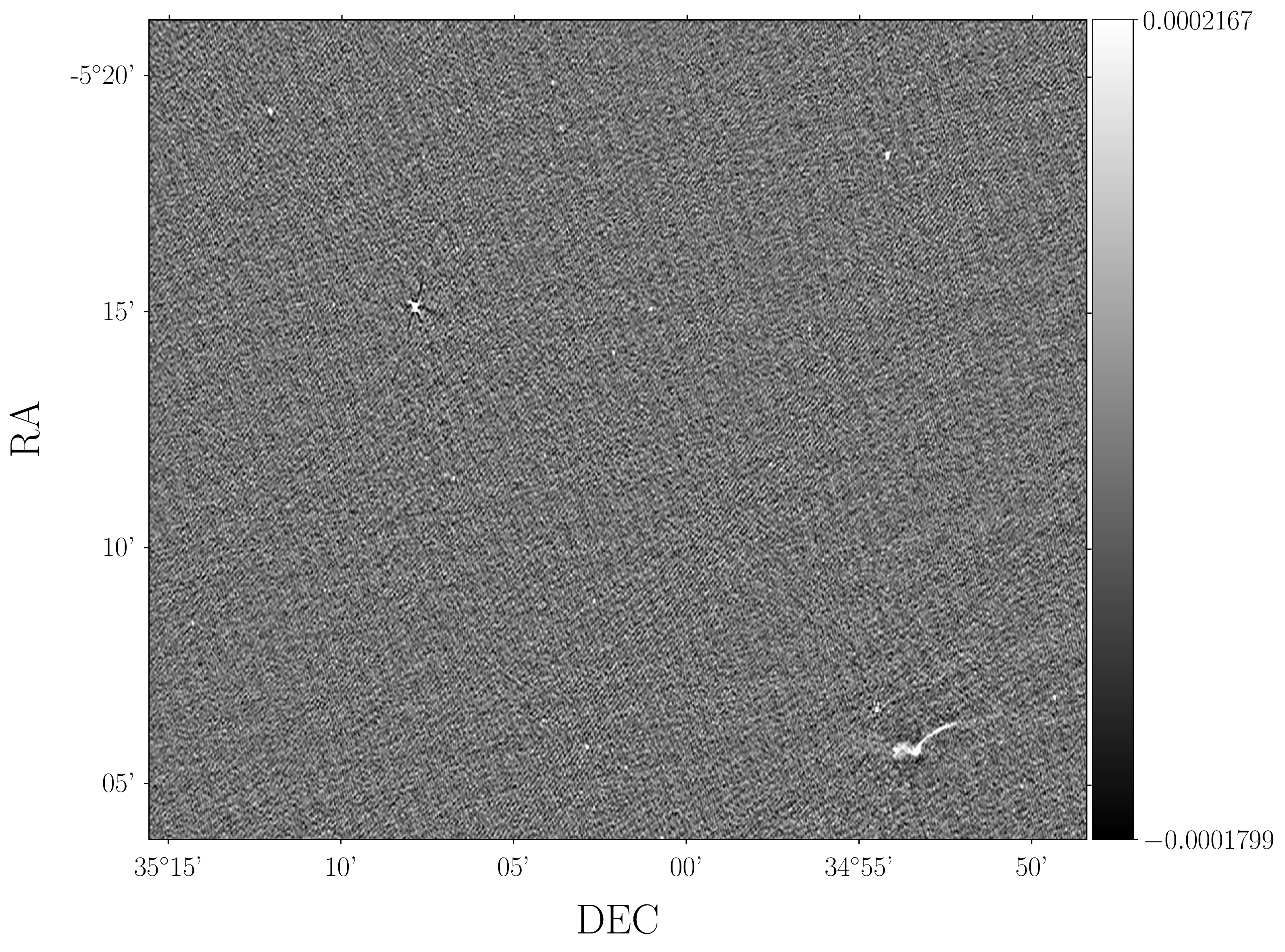} %deep_robust_fint_128_tint_1_image.pdf
\caption{Robust solver, 9 sec intervals}\label{robust_tint1}
 \end{subfigure}
\begin{subfigure}{.48\textwidth}
 \includegraphics[width=8cm,height=6.5cm,keepaspectratio]{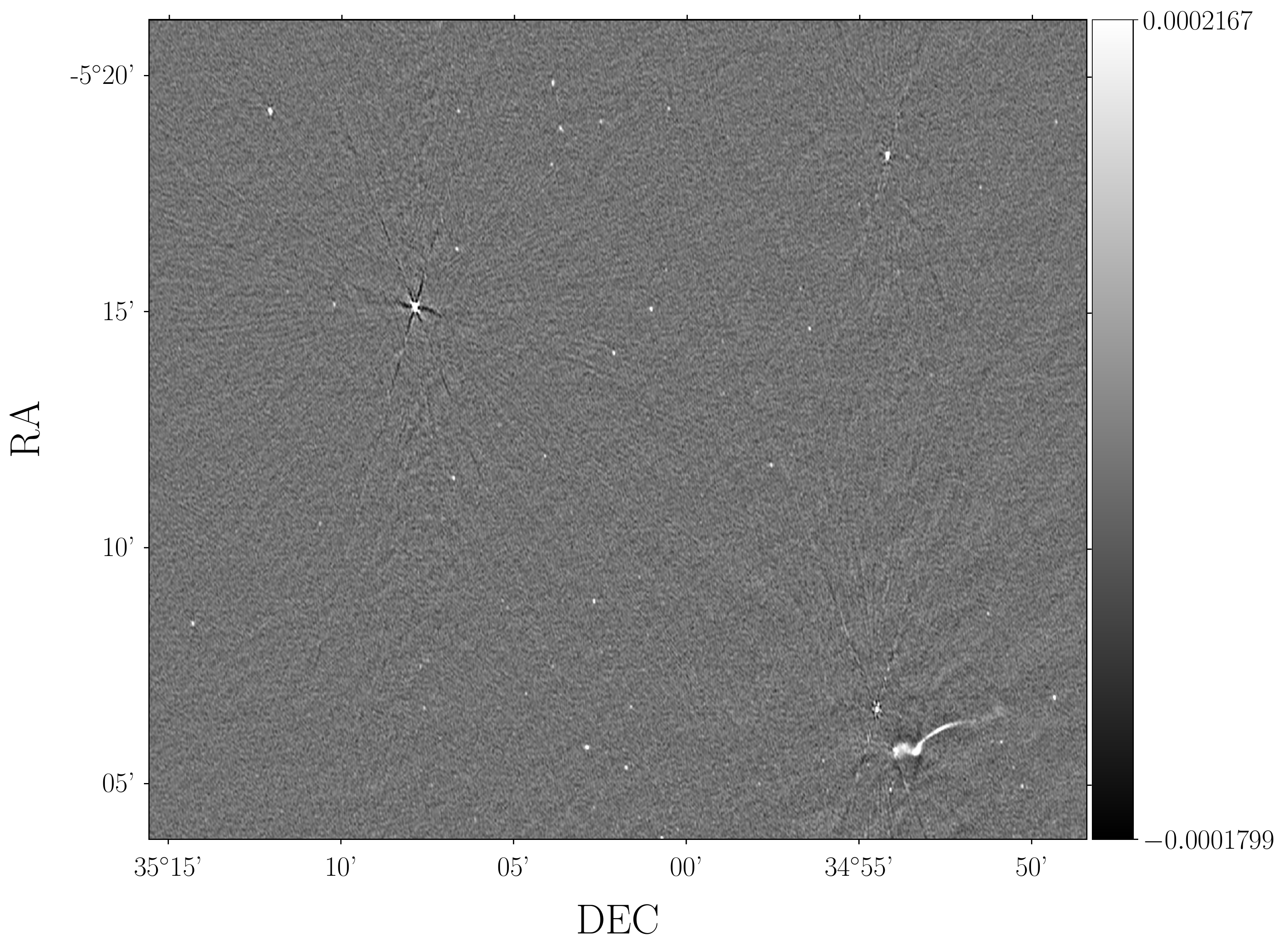} %deep_complex_fint_128_tint_64_image.pdf
\caption{Complex solver, 7.5 min intervals} \label{complex_tint64}
 \end{subfigure}
% \vspace{-1cm}
\begin{subfigure}{.48\textwidth}
  \includegraphics[width=8cm,height=6.5cm,keepaspectratio]{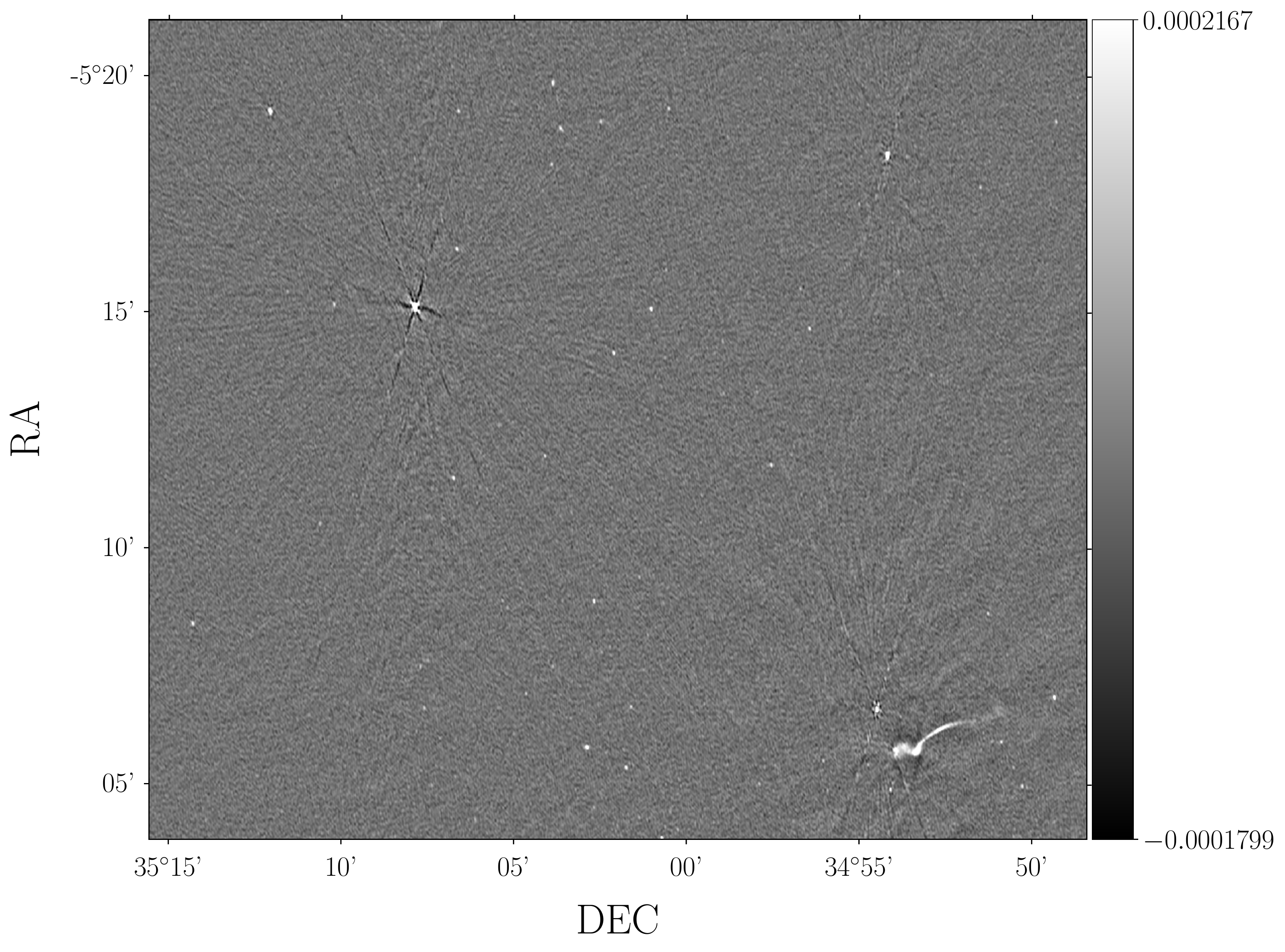} %deep_robust_fint_128_tint_64_image.pdf
\caption{Robust solver, 7.5 min intervals}\label{robust_tint64}
 \end{subfigure}
\begin{subfigure}{.48\textwidth}
  \includegraphics[width=8cm,height=6.5cm,keepaspectratio]{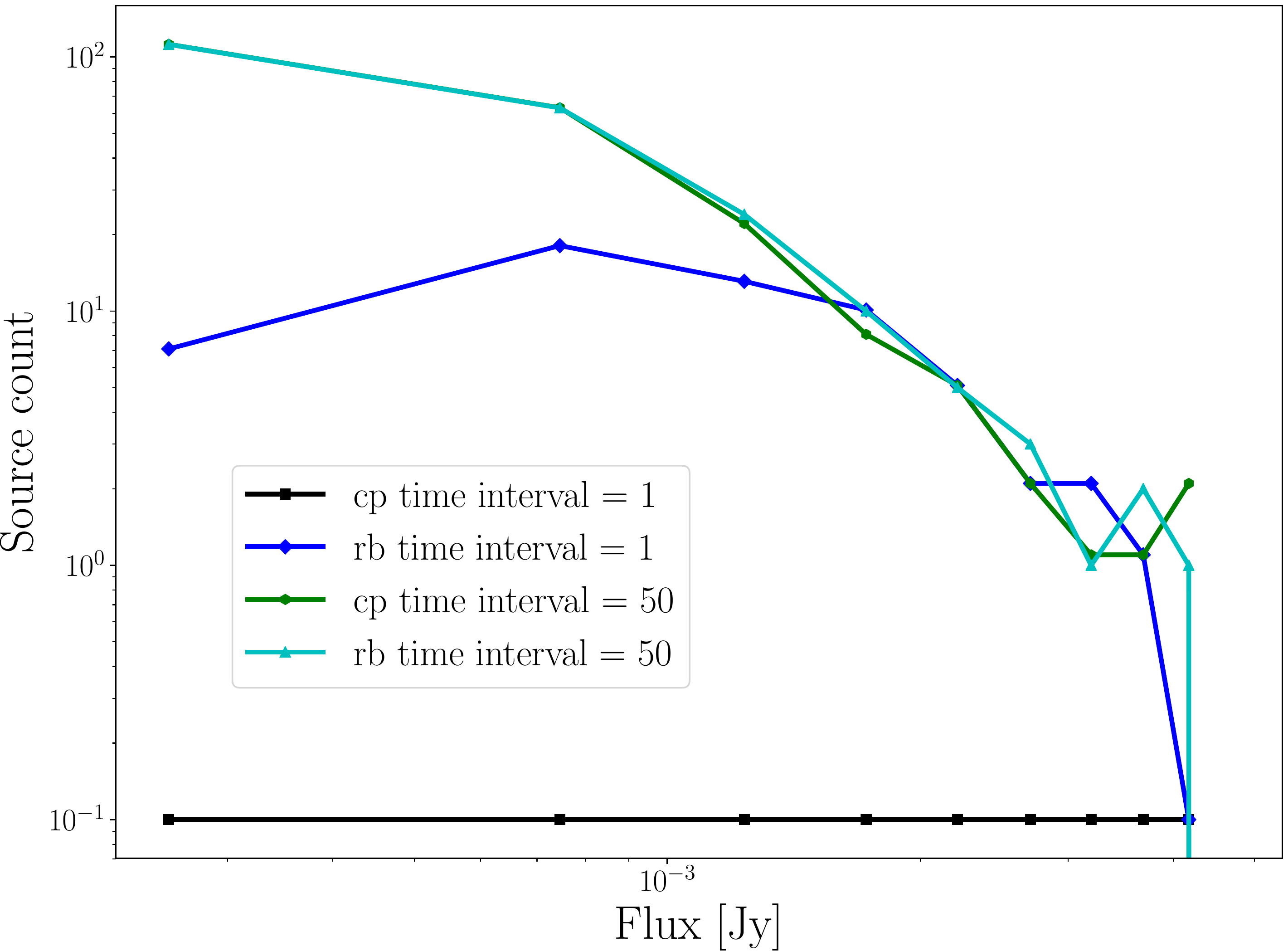} %deep2_tint_1_count.pdf
\caption{Source counts}\label{source_count}
 \end{subfigure}
\caption{An image of the centre of the VIDEO field before and after calibration with different solution intervals: (a) before self-calibration; (b) after calibration using the complex solver with a time interval of 9 secs. The image gets worse because RFI contaminates some gain solutions; (c) after calibration using the robust solver with a time of interval 9 secs. Most artefacts from the uncalibrated image are gone, but the noise level is increased due to the low SNR of the solutions; (d) after calibration using the complex solver with a time interval of 7.5 mins. The RFI-induced artefacts are gone because they have been averaged out by the long solution interval; (e) after calibration using the robust solver with a time interval of 7.5 mins. (f) Source counts, showing no detections for the complex solver-calibrated image with a time interval of 9 secs.}
\label{DEEP field images}
\end{figure*}

\begin{figure*}
\begin{subfigure}{.48\textwidth}
  \includegraphics[width = 0.98\linewidth]{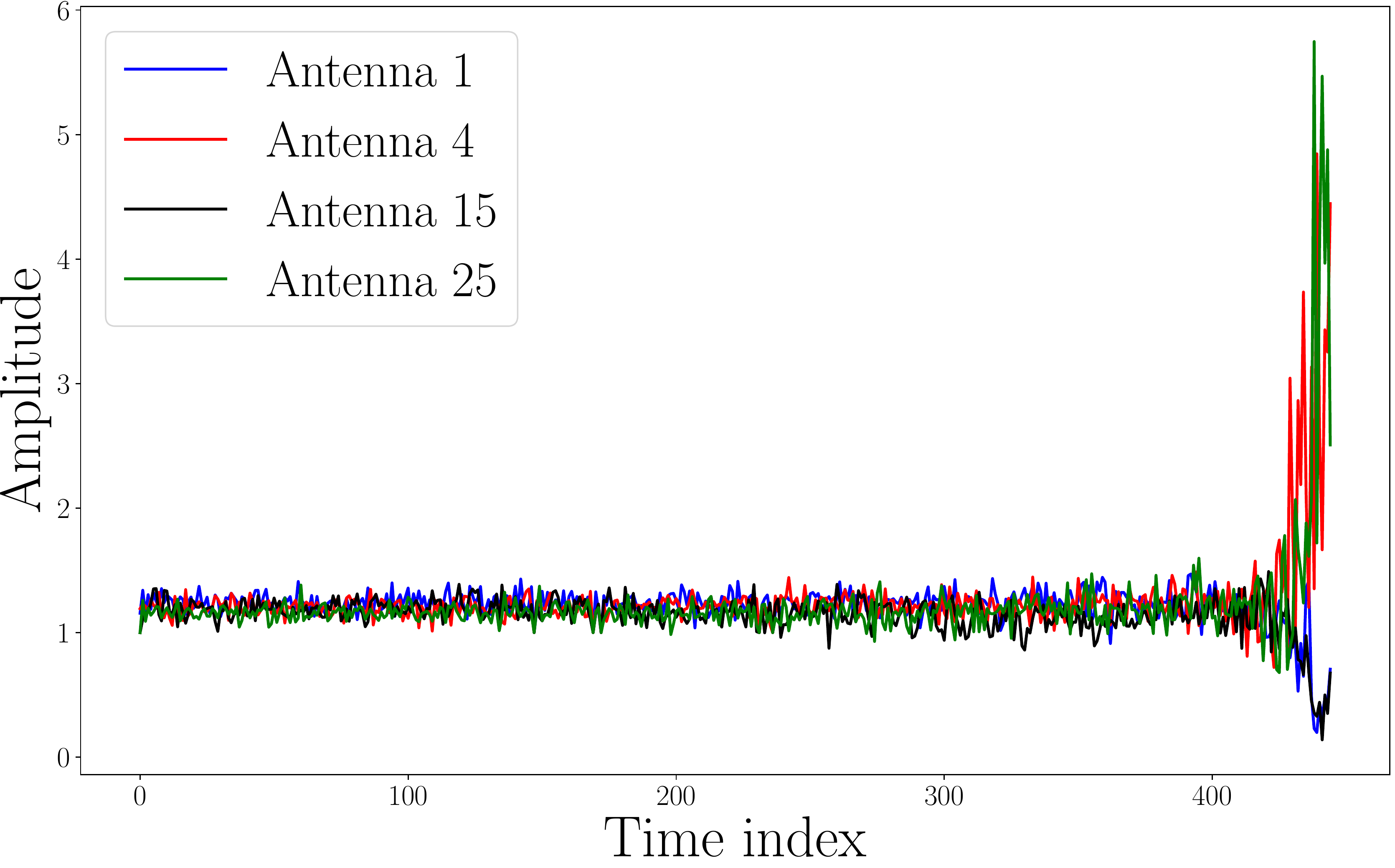} %complex_gains_fint_128_tint_1_ant_1_amp.pdf
  \caption{complex gains time-int = 9 secs}\label{complex_gains1}
 \end{subfigure}
\begin{subfigure}{.48\textwidth}
  \includegraphics[width = 0.98\linewidth]{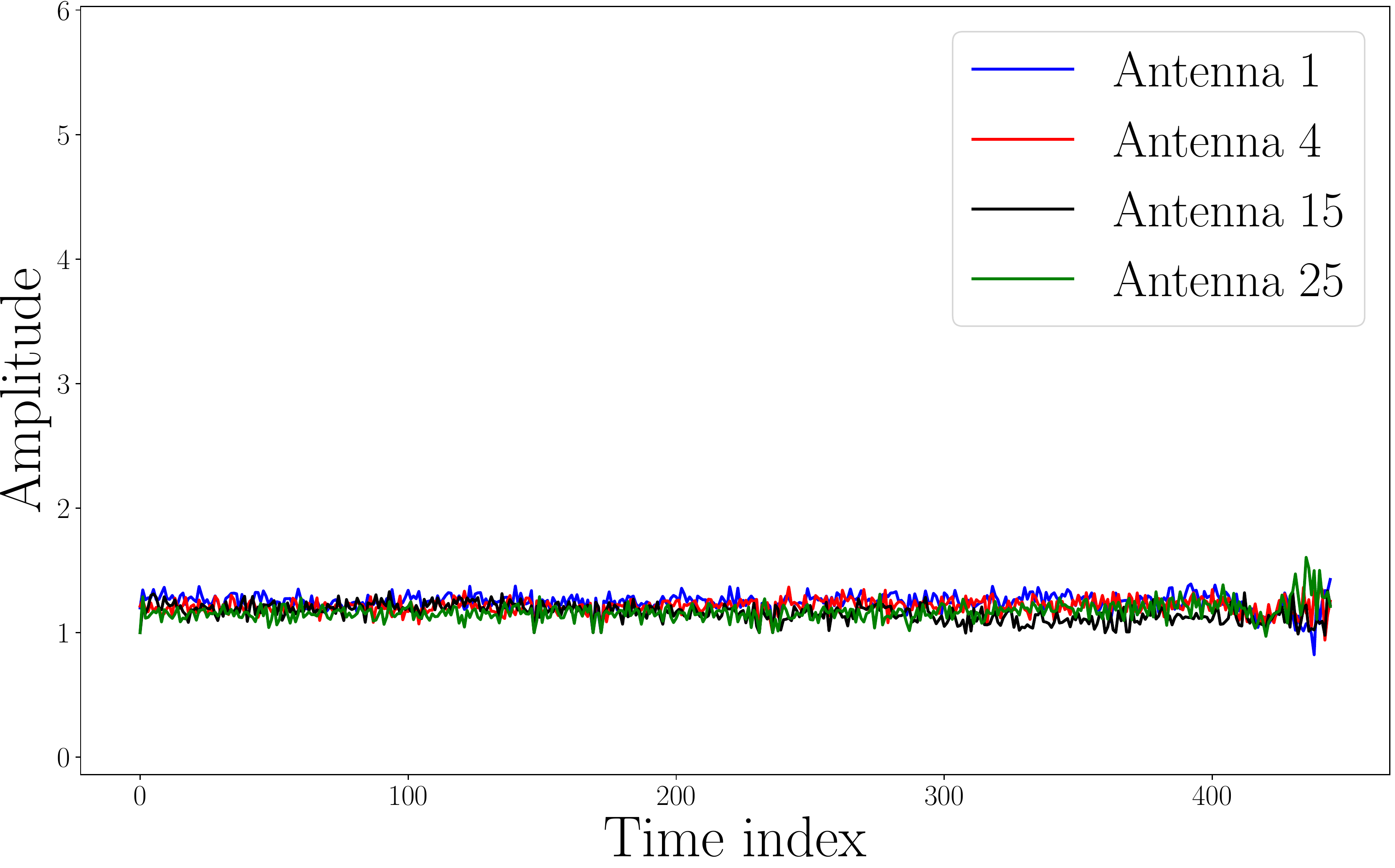} %robust_gains_fint_128_tint_1_ant_1_amp.pdf
\caption{robust gains time-int = 9 secs}\label{robust_gains1}
 \end{subfigure}
 \begin{subfigure}{.48\textwidth}
  \includegraphics[width = 0.98\linewidth]{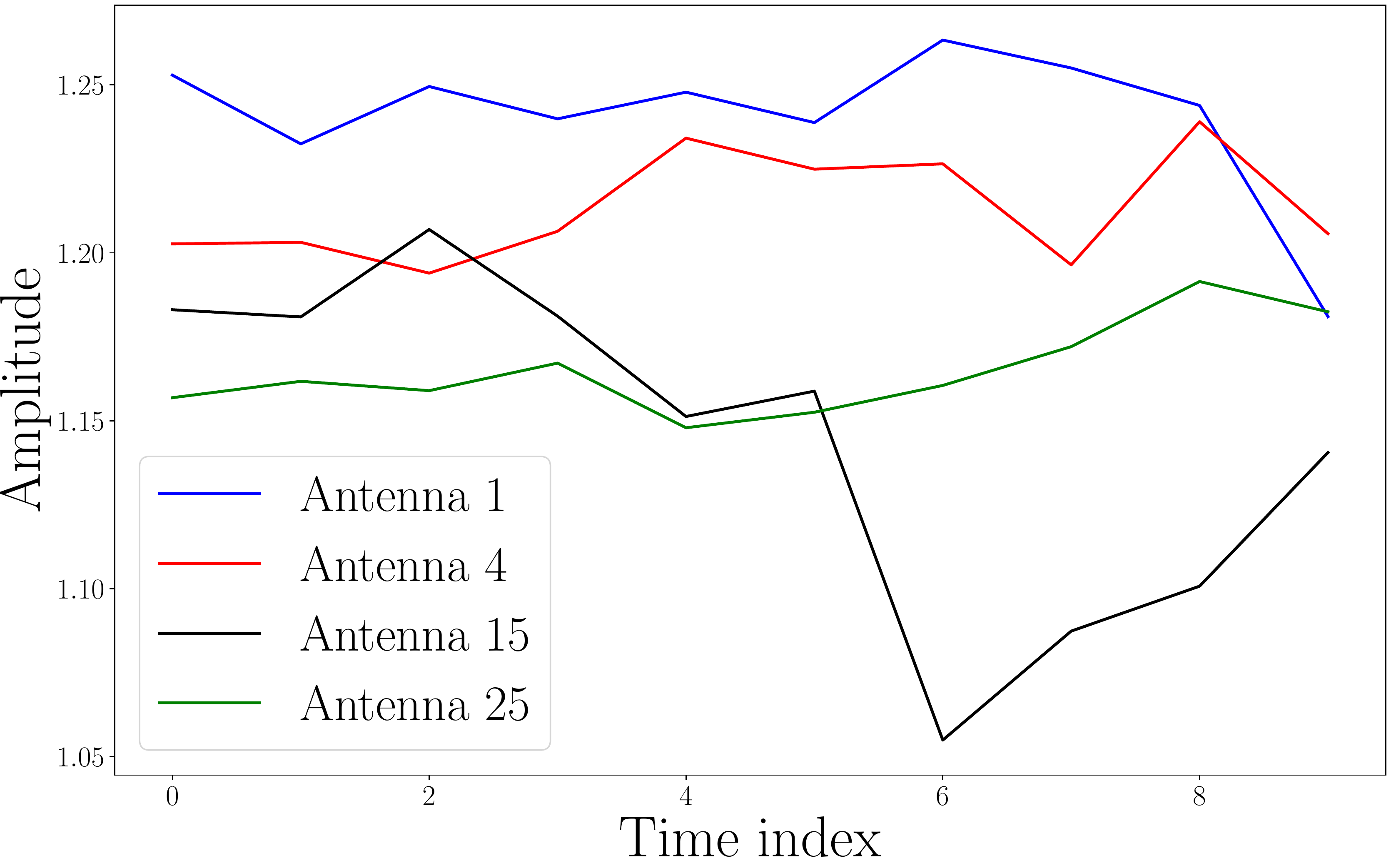} %complex_gains_fint_128_tint_64_ant_1_amp.pdf
  \caption{complex gains time-int = 7.5 mins}\label{complex_gains2}
 \end{subfigure}
\begin{subfigure}{.48\textwidth}
  \includegraphics[width = 0.98\linewidth]{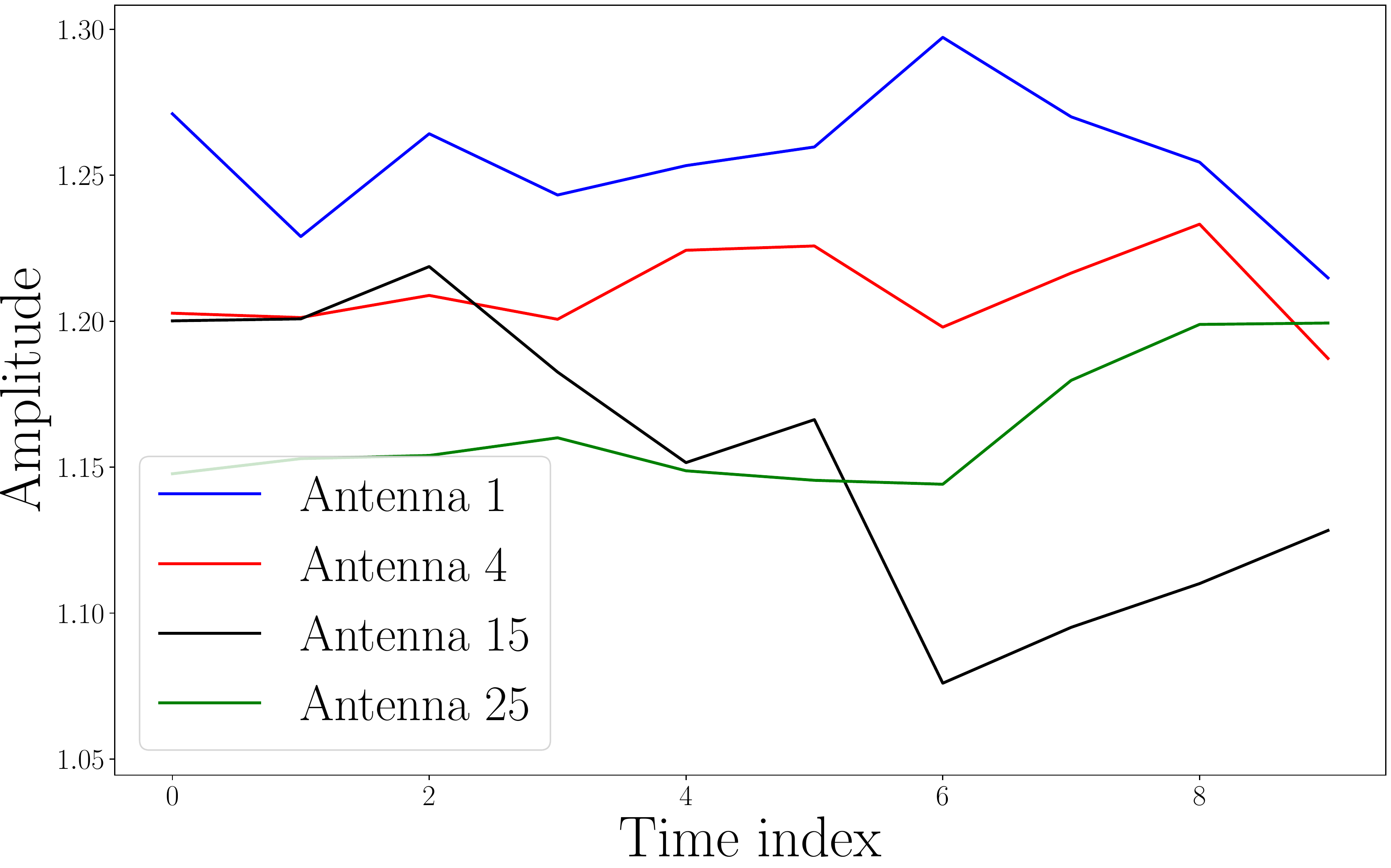} %robust_gains_fint_128_tint_64_ant_1_amp.pdf
\caption{robust gains time-int = 7.5 mins}\label{robust_gains2}
 \end{subfigure}
\caption{The amplitudes of the estimated gains for both solvers with time intervals 9 secs and 7.5 mins. For the complex solver (a) we can see various peaks which are absent for the robust solver (b). These peaks appear exactly at the times where RFI dominates the visibilities (see Figure \ref{deep_2_rfi}). The robust solver highly attenuates these peaks as a result of the weighting. (c) and (d) show that, with a large time interval of 7.5 mins, the peaks are average out for the complex solver.}
\label{deep_gains}
\end{figure*}

Images of the 1GC-corrected and post-2GC data are shown in Figure \ref{DEEP field images}. The strong artefacts present in Figure~\ref{deep_befor}, which are not visible in the simulated data (see Figure \ref{deep_RFI_sim}), are caused by the primary beam: as the earth rotates during an observation, the sources move through the beam and produce these artefacts. 
Figure \ref{complex_tint1} and \ref{robust_tint1} show that the robust solver removes the artefacts in the data (see Figure \ref{deep_befor}), whereas the solutions from the complex solver are similar to the predictions of the RFI simulation. On the other hand, at a time interval of 7.5 mins, both solvers produce good results and the artefacts are effectively removed. A look at the gain plots in Figure \ref{deep_gains} provides additional insight. At a time interval of 9 secs, the RFI occupancy in some of the time intervals is rather high, leading to biased gain solutions derived from the complex solver. These biased solutions propagate errors into the corrected visibilities, resulting in strong imaging  artefacts. The robust solver does not suffer from this effect because very low weights are applied to these visibilities, thus effectively flagging them during the computation of the gains. We observe more noise in Figure~\ref{robust_tint1}: at low SNR, if short time/frequency intervals are used for calibration, the solver fits noise instead of signal and this may result in an increase in the noise level of the corrected data. We leave the investigation of this subtle trade-off between solution interval width and SNR for future research. Note that as the solution interval increases, the performance of both solvers converges. The RFI contribution is averaged out by the long time interval, so the complex solver is able to perform adequately, as shown by the output gains plot (Fig.~\ref{deep_gains}). 

Finally, we conclude by presenting Fig. \ref{source_count} which shows a (log scale) plot of the source counts extracted from the different images. Figure \ref{source_count} gives us an insight into how such low-level RFI could affect our science. In particular, we don't detect any sources in the calibrated image when we use the complex solver with a time interval of 9 secs. Therefore, low-level RFI needs to be handled properly during calibration, even when it is not immediately obvious in the image domain. This final point is particularly relevant for mJy and $\mu$Jy science targets. 
\section{Conclusions}
Accurate, efficient and fast data processing will be essential for the new generation of radio telescopes such as the MeerKAT and the upcoming Square Kilometre Array (SKA). In this paper, we describe how to mitigate the effects of unmodelled sources and RFI during calibration. Leveraging recent advancements in the field of complex optimisation, we implement a robust calibration algorithm based on a CST, inspired by \citet{kazemi2013robust}. The code for the newly implemented robust solver is publicly available, and has been integrated into the recently released CubiCal package \citep{kenyon2018cubical}.

We apply this solver to both simulated and real data sets and obtain significant improvements in calibration scenarios involving incomplete sky models and RFI-contaminated data. Statistical analysis of RI visibilities shows that unmodelled point sources do not change the underlying distribution of residual visibilities from Gaussian to a Student's t but do increase the perceived variance of the residuals, thereby reducing the SNR during calibration. The robust solver, which employs an iterative weighting scheme, mitigates this effect during calibration by assigning small weights to visibilities with residuals far away from the estimated data covariance. The weighting improves the calibration solutions and reduces the amount of flux suppressed as a result of the unmodelled sources.

The same concept applies when calibrating data containing low-level unflagged RFI. The robust solver mitigates RFI-contaminated visibilities by assigning them smaller weights, hence preventing the calibration solutions from blowing up and propagating the RFI into the corrected visibilities. For both the incomplete sky models and RFI scenarios, we also observe that using long solution intervals can also mitigate this effect.

As far as future work is concerned, it will be particularly interesting to see how the robust solver performs on extended sources and diffuse emission, since the visibility contribution of these is restricted to shorter baselines. Proper benchmarking needs to performed in order to identify the optimal settings for the robust solver, and  the different conditions under which it improves calibration. The fact that the robust-$\mathbf{I}$ solver works remarkably well in the high SNR simulation (where the true covariance of the residuals is much higher than $\mathbf{I}$) suggests that scaling down the covariance could improve results of the robust solver. We intend to investigate this in a future work. The algorithm we implemented uses a non Gaussian likelihood function, but is entirely independent of the RIME model thus extending it to other CubiCal solvers should be a straightforward process.

The flux suppression results (particularly, those in  Figure~\ref{di_supp_snr}) show some surprising trends. The strong link between model concentration and flux suppression has not, to our knowledge, been appreciated before. 
Why do we observe markedly more flux suppression with the traditional solver when calibrating with a highly dispersed sky model? Answering this question necessitates revisiting and extending the calibration artefact studies of \citet{grobler2014calibration} and \citet{wijnholds2016calibration}.

\section*{Acknowledgements}
This research is supported by the South African Research Chairs Initiative of the Department of Science and Technology and National Research Foundation, the Royal Socienty and the Newton Fund (grant NA150184), and the South African Radio Astronomy Observatory (SARAO).

%%%%%%%%%%%%%%%%%%%%%%%%%%%%%%%%%%%%%%%%%%%%%%%%%%
%%%%%%%%%%%%%%%%%%%% REFERENCES %%%%%%%%%%%%%%%%%%
% The best way to enter references is to use BibTeX:
\bibliographystyle{mnras}
\bibliography{Bibliography} % if your bibtex file is called example.bib

\begin{thebibliography}{}
\makeatletter
\relax
\def\mn@urlcharsother{\let\do\@makeother \do\$\do\&\do\#\do\^\do\_\do\%\do\~}
\def\mn@doi{\begingroup\mn@urlcharsother \@ifnextchar [ {\mn@doi@}
  {\mn@doi@[]}}
\def\mn@doi@[#1]#2{\def\@tempa{#1}\ifx\@tempa\@empty \href
  {http://dx.doi.org/#2} {doi:#2}\else \href {http://dx.doi.org/#2} {#1}\fi
  \endgroup}
\def\mn@eprint#1#2{\mn@eprint@#1:#2::\@nil}
\def\mn@eprint@arXiv#1{\href {http://arxiv.org/abs/#1} {{\tt arXiv:#1}}}
\def\mn@eprint@dblp#1{\href {http://dblp.uni-trier.de/rec/bibtex/#1.xml}
  {dblp:#1}}
\def\mn@eprint@#1:#2:#3:#4\@nil{\def\@tempa {#1}\def\@tempb {#2}\def\@tempc
  {#3}\ifx \@tempc \@empty \let \@tempc \@tempb \let \@tempb \@tempa \fi \ifx
  \@tempb \@empty \def\@tempb {arXiv}\fi \@ifundefined
  {mn@eprint@\@tempb}{\@tempb:\@tempc}{\expandafter \expandafter \csname
  mn@eprint@\@tempb\endcsname \expandafter{\@tempc}}}

\bibitem[\protect\citeauthoryear{Bishop}{Bishop}{2006}]{bishop2006pattern}
Bishop C.~M.,  2006, Pattern Recognition and Machine Learning (Information
  Science and Statistics).
Springer-Verlag, Berlin, Heidelberg

\bibitem[\protect\citeauthoryear{Bonnassieux, Tasse, Smirnov  \&
  Zarka}{Bonnassieux et~al.}{2018}]{bonnassieux2018variance}
Bonnassieux E.,  Tasse C.,  Smirnov O.,   Zarka P.,  2018, Astronomy \&
  Astrophysics, 615, A66

\bibitem[\protect\citeauthoryear{{Cornwell} \& {Wilkinson}}{{Cornwell} \&
  {Wilkinson}}{1981}]{cornwell1981wilkinson}
{Cornwell} T.~J.,  {Wilkinson} P.~N.,  1981, \mn@doi [\mnras]
  {10.1093/mnras/196.4.1067}, \href
  {http://adsabs.harvard.edu/abs/1981MNRAS.196.1067C} {196, 1067}

\bibitem[\protect\citeauthoryear{Grobler, Nunhokee, Smirnov, van Zyl  \& de
  Bruyn}{Grobler et~al.}{2014}]{grobler2014calibration}
Grobler T.~L.,  Nunhokee C.~D.,  Smirnov O.~M.,  van Zyl A.~J.,   de Bruyn
  A.~G.,  2014, \mn@doi [Monthly Notices of the Royal Astronomical Society]
  {10.1093/mnras/stu268}, 439, 4030

\bibitem[\protect\citeauthoryear{Grobler, Stewart, Wijnholds, Kenyon  \&
  Smirnov}{Grobler et~al.}{2016}]{grobler2016calibration}
Grobler T.,  Stewart A.,  Wijnholds S.,  Kenyon J.,   Smirnov O.,  2016,
  Monthly Notices of the Royal Astronomical Society, 461, 2975

\bibitem[\protect\citeauthoryear{{Grobler}, {Bernardi}, {Kenyon}, {Parsons}  \&
  {Smirnov}}{{Grobler} et~al.}{2018}]{grobler2018redundant}
{Grobler} T.~L.,  {Bernardi} G.,  {Kenyon} J.~S.,  {Parsons} A.~R.,   {Smirnov}
  O.~M.,  2018, \mn@doi [\mnras] {10.1093/mnras/sty357}, \href
  {http://adsabs.harvard.edu/abs/2018MNRAS.476.2410G} {476, 2410}

\bibitem[\protect\citeauthoryear{{Hamaker}, {Bregman}  \& {Sault}}{{Hamaker}
  et~al.}{1996}]{hamaker1996understanding}
{Hamaker} J.~P.,  {Bregman} J.~D.,   {Sault} R.~J.,  1996, \aaps, \href
  {http://adsabs.harvard.edu/abs/1996A%26AS..117..137H} {117, 137}

\bibitem[\protect\citeauthoryear{{Heywood}, {Hale}, {Jarvis}, {Makhathini},
  {Peters}, {Sebokolodi}  \& {Smirnov}}{{Heywood} et~al.}{submitted}]{IanVIDEO}
{Heywood} I.,  {Hale} C.~L.,  {Jarvis} M. J.,  {Makhathini} S.,  {Peters} J.,
  {Sebokolodi} M.,   {Smirnov} O.~M.,  submitted, Monthly Notices of the Royal
  Astronomical Society

\bibitem[\protect\citeauthoryear{Jonas \& Team}{Jonas \&
  Team}{2018}]{jonas2018meerkat}
Jonas J.,  Team M.,  2018, MeerKAT Science: On the Pathway to the SKA
  (MeerKAT2016), eds Taylor P, Camilo F, Leeuw L, Moodley K (International
  School for Advanced Studies, Trieste, Italy), 277

\bibitem[\protect\citeauthoryear{{Kazemi} \& {Yatawatta}}{{Kazemi} \&
  {Yatawatta}}{2013}]{kazemi2013robust}
{Kazemi} S.,  {Yatawatta} S.,  2013, \mn@doi [\mnras] {10.1093/mnras/stt1347},
  \href {http://adsabs.harvard.edu/abs/2013MNRAS.435..597K} {435, 597}

\bibitem[\protect\citeauthoryear{Kenyon}{Kenyon}{2019}]{kenyon2019link}
Kenyon J.,  2019, PhD thesis, Rhodes University, \url
  {http://hdl.handle.net/10962/92341}

\bibitem[\protect\citeauthoryear{{Kenyon}, {Smirnov}, {Grobler}  \&
  {Perkins}}{{Kenyon} et~al.}{2018}]{kenyon2018cubical}
{Kenyon} J.~S.,  {Smirnov} O.~M.,  {Grobler} T.~L.,   {Perkins} S.~J.,  2018,
  \mn@doi [\mnras] {10.1093/mnras/sty1221}, \href
  {http://adsabs.harvard.edu/abs/2018MNRAS.478.2399K} {478, 2399}

\bibitem[\protect\citeauthoryear{{Kreutz-Delgado}}{{Kreutz-Delgado}}{2009}]{kreutzdelgado2009}
{Kreutz-Delgado} K.,  2009, preprint, \href
  {http://adsabs.harvard.edu/abs/2009arXiv0906.4835K} {} (\mn@eprint {arXiv}
  {0906.4835})

\bibitem[\protect\citeauthoryear{Lange, Little  \& Taylor}{Lange
  et~al.}{1989}]{lange1989robust}
Lange K.~L.,  Little R.~J.,   Taylor J.~M.,  1989, Journal of the American
  Statistical Association, 84, 881

\bibitem[\protect\citeauthoryear{Madsen, Nielsen  \& Tingleff}{Madsen
  et~al.}{2004}]{madsen2004methods}
Madsen K.,  Nielsen H.,   Tingleff O.,  2004, Informatics and Mathematical
  Modelling Technical University of Denmark, p.~60

\bibitem[\protect\citeauthoryear{{Mohan} \& {Rafferty}}{{Mohan} \&
  {Rafferty}}{2015}]{mohan2015pybdsf}
{Mohan} N.,  {Rafferty} D.,  2015, {PyBDSF: Python Blob Detection and Source
  Finder}, Astrophysics Source Code Library (\mn@eprint {ascl} {1502.007})

\bibitem[\protect\citeauthoryear{{Noordam} \& {Smirnov}}{{Noordam} \&
  {Smirnov}}{2010}]{noordam2010meqtrees}
{Noordam} J.~E.,  {Smirnov} O.~M.,  2010, \mn@doi [\aap]
  {10.1051/0004-6361/201015013}, \href
  {http://adsabs.harvard.edu/abs/2010A%26A...524A..61N} {524, A61}

\bibitem[\protect\citeauthoryear{Nunhokee}{Nunhokee}{2015}]{nunhokee2015link}
Nunhokee C.~D.,  2015, Master's thesis, Rhodes University, \url
  {http://hdl.handle.net/10962/d1017900}

\bibitem[\protect\citeauthoryear{{Offringa}, {van de Gronde}  \&
  {Roerdink}}{{Offringa} et~al.}{2012}]{offringa2012morphological}
{Offringa} A.~R.,  {van de Gronde} J.~J.,   {Roerdink} J.~B.~T.~M.,  2012,
  \mn@doi [\aap] {10.1051/0004-6361/201118497}, \href
  {http://adsabs.harvard.edu/abs/2012A%26A...539A..95O} {539, A95}

\bibitem[\protect\citeauthoryear{Offringa et~al.,}{Offringa
  et~al.}{2014}]{offringa2014wsclean}
Offringa A.,  et~al., 2014, Monthly Notices of the Royal Astronomical Society,
  444, 606

\bibitem[\protect\citeauthoryear{Offringa et~al.,}{Offringa
  et~al.}{2015}]{offringa2015}
Offringa A.~R.,  et~al., 2015, \mn@doi [Publications of the Astronomical
  Society of Australia] {10.1017/pasa.2015.7}, 32, e008

\bibitem[\protect\citeauthoryear{{Ollier}, {Korso}, {Boyer}, {Larzabal}  \&
  {Pesavento}}{{Ollier} et~al.}{2017}]{ollier2017robust}
{Ollier} V.,  {Korso} M.~N.~E.,  {Boyer} R.,  {Larzabal} P.,   {Pesavento} M.,
  2017, \mn@doi [IEEE Transactions on Signal Processing]
  {10.1109/TSP.2017.2733496}, \href
  {http://adsabs.harvard.edu/abs/2017ITSP...65.5649O} {65, 5649}

\bibitem[\protect\citeauthoryear{{Perkins}, {Marais}, {Zwart}, {Natarajan},
  {Tasse}  \& {Smirnov}}{{Perkins} et~al.}{2015}]{perkins2015}
{Perkins} S.~J.,  {Marais} P.~C.,  {Zwart} J.~T.~L.,  {Natarajan} I.,  {Tasse}
  C.,   {Smirnov} O.,  2015, \mn@doi [Astronomy and Computing]
  {10.1016/j.ascom.2015.06.003}, \href
  {http://adsabs.harvard.edu/abs/2015A%26C....12...73P} {12, 73}

\bibitem[\protect\citeauthoryear{{Rasmussen} \& {Williams}}{{Rasmussen} \&
  {Williams}}{2006}]{rasmussen2006gaussian}
{Rasmussen} C.~E.,  {Williams} C.~K.~I.,  2006, {Gaussian Processes for Machine
  Learning}.
Springer

\bibitem[\protect\citeauthoryear{{Salvini} \& {Wijnholds}}{{Salvini} \&
  {Wijnholds}}{2014}]{salvini2014fast}
{Salvini} S.,  {Wijnholds} S.~J.,  2014, \mn@doi [\aap]
  {10.1051/0004-6361/201424487}, \href
  {http://adsabs.harvard.edu/abs/2014A%26A...571A..97S} {571, A97}

\bibitem[\protect\citeauthoryear{Schwab}{Schwab}{1981}]{schwab1981vla}
Schwab F.,  1981, available online through http://www. vla. nrao.
  edu/memos/sci/136. pdf

\bibitem[\protect\citeauthoryear{Smirnov}{Smirnov}{2011}]{smirnov2011revisiting}
Smirnov O.~M.,  2011, Astronomy \& Astrophysics, 527, A106

\bibitem[\protect\citeauthoryear{{Smirnov} \& {Tasse}}{{Smirnov} \&
  {Tasse}}{2015}]{smirnov2015radio}
{Smirnov} O.~M.,  {Tasse} C.,  2015, \mn@doi [\mnras] {10.1093/mnras/stv418},
  \href {http://adsabs.harvard.edu/abs/2015MNRAS.449.2668S} {449, 2668}

\bibitem[\protect\citeauthoryear{Sorber, van Barel  \& de Lathauwer}{Sorber
  et~al.}{2012}]{sorber2012}
Sorber L.,  van Barel M.,   de Lathauwer L.,  2012, \mn@doi [SIAM J. Optim.]
  {10.1137/110832124}, 22, 879

\bibitem[\protect\citeauthoryear{{Tasse}}{{Tasse}}{2014a}]{tasse2014}
{Tasse} C.,  2014a, preprint, \href
  {http://adsabs.harvard.edu/abs/2014arXiv1410.8706T} {} (\mn@eprint {arXiv}
  {1410.8706})

\bibitem[\protect\citeauthoryear{{Tasse}}{{Tasse}}{2014b}]{tasse2014nonlinear}
{Tasse} C.,  2014b, \mn@doi [\aap] {10.1051/0004-6361/201423503}, \href
  {http://adsabs.harvard.edu/abs/2014A%26A...566A.127T} {566, A127}

\bibitem[\protect\citeauthoryear{Wijnholds, Smirnov  \& Grobler}{Wijnholds
  et~al.}{2016}]{wijnholds2016calibration}
Wijnholds S.~J.,  Smirnov O.~M.,   Grobler T.~L.,  2016, \mn@doi [Monthly
  Notices of the Royal Astronomical Society] {10.1093/mnras/stw118}, 457, 2331

\bibitem[\protect\citeauthoryear{{Wirtinger}}{{Wirtinger}}{1927}]{wirtinger1927}
{Wirtinger} W.,  1927, \mn@doi [Mathematische Annalen] {10.1007/BF01447872},
  97, 357

\bibitem[\protect\citeauthoryear{Yatawatta}{Yatawatta}{2013}]{yatawatta2013radio}
Yatawatta S.,  2013, in 2013 IEEE International Conference on Acoustics, Speech
  and Signal Processing. pp 3866--3870

\bibitem[\protect\citeauthoryear{Yatawatta}{Yatawatta}{2015}]{10.1093/mnras/stv596}
Yatawatta S.,  2015, \mn@doi [Monthly Notices of the Royal Astronomical
  Society] {10.1093/mnras/stv596}, 449, 4506

\bibitem[\protect\citeauthoryear{Yatawatta, De~Clercq, Spreeuw  \&
  Diblen}{Yatawatta et~al.}{2019}]{yatawatta2019stochastic}
Yatawatta S.,  De~Clercq L.,  Spreeuw H.,   Diblen F.,  2019, arXiv preprint
  arXiv:1904.05619

\makeatother
\end{thebibliography}
% Alternatively you could enter them by hand, like this:
% This method is tedious and prone to error if you have lots of references
% \begin{thebibliography}{99}
% \bibitem[\protect\citeauthoryear{Author}{2012}]{Author2012}
% Author A.~N., 2013, Journal of Improbable Astronomy, 1, 1
% \bibitem[\protect\citeauthoryear{Others}{2013}]{Others2013}
% Others S., 2012, Journal of Interesting Stuff, 17, 198
% \end{thebibliography}
%%%%%%%%%%%%%%%%%%%%%%%%%%%%%%%%%%%%%%%%%%%%%%%%%%
%%%%%%%%%%%%%%%%% APPENDICES %%%%%%%%%%%%%%%%%%%%%
\appendix
\section{Expectation Maximisation for Non-linear Models with Proper CST Noise}
\label{sec:appendix}
Suppose we have a measurement model given by
\begin{equation}
\By = f(\Bx) + \epsilon, \quad \mbox{where} \quad \By \in \mathbb{C}^D, ~ \Bx \in \mathbb{C}^M, ~ \epsilon \sim \CST{\By| \Bmu=f(\Bx), \BLambda, v}
\end{equation}
where $f: \mathbb{C}^M \rightarrow \mathbb{C}^D$ is some non-linear function and we want to find the maximum likelihood (ML) estimate of $\Bx$ given a set of data $Y$. The ML solution requires solving for the parameters $\Bx$ as well as the parameters defining the CST distribution i.e. $v$ and $\BLambda$. We will denote these as a single parameter vector, $\theta$. Unfortunately, the CST is not part of the exponential family and the log-likelihood is generally clumsy to work with. We will now illustrate how the ML solution can be obtained using an iteratively reweighted complex NLLS algorithm. 

From Section \ref{sec:complex_t_cal}, it is clear that we can view the distribution of each data point $\By_i$ as an infinite mixture of proper complex normal distributions with variance drawn from a Gamma distribution i.e.
\begin{equation}
P(\By_i | \theta) = \int_0^\infty \CN\left(\By_i|\Bmu_i=f(\Bx), (\tau_i \BLambda)^{-1}\right) \Gam{\tau_i| v, v} d \tau_i,
\label{marginal}
\end{equation} 
where we have left the dependence of $f$ implicit for notational simplicity. In this case, assuming conditional independence of the data, the likelihood given $N$ data points is simply
\begin{align}
P(Y | \theta) &= \prod_{i=1}^{N} P(\By_i | \theta), \nonumber \\
P(Y | \theta) &= \int dZ \prod_{i=1}^{N} \CN\left(\By_i|\Bmu_i, (\tau_i \BLambda)^{-1}\right) \Gam{\tau_i| v, v}
\label{fullmarginal}
\end{align}
where we have denoted $Z = [\tau_1, \tau_2, \cdots, \tau_N]^T$ as the set of latent variables corresponding to the scale parameter for each data point and used the fact that all the $\tau_i$ are independent to exchange the order of the product and the integral. \\

The form \eqref{fullmarginal} can now be solved using using the expectation maximisation (EM) algorithm. For notational convenience we denote the joint density (i.e. the integrand of \eqref{fullmarginal}) by $P(Y, Z|\theta)$. The key idea behind EM is to identify latent variables Z, governed by a distribution $q(Z)$ for example, for which the joint density $P(Y, Z| \theta)$ is easier to evaluate than the marginal in \eqref{fullmarginal}. The trick is then to decompose the log of the marginal density into two functionals viz.
\begin{align}
\log P(Y| \theta) &= \int dZ q(Z) \log\left( \frac{P(Y, Z|\theta)}{q(Z)} \right) - \int dZ q(Z) \log\left(\frac{P(Z|Y, \theta)}{q(Z)}\right), \\
&= \mathcal{L}(q(Z), \theta) + KL\left(q(Z)\|P(Z|Y,\theta)\right).
\end{align} 
Noting that the last term is the Kullback-Leibler divergence which satisfies $KL\left(q(Z)\|P(Z|Y,\theta)\right) \geq 0$ with equality holding iff $q(z) = P(Z|Y,\theta)$, we see that $\mathcal{L}(q(Z), \Bx)$ is a lower bound on $\log P(Y| \theta)$. This implies that the optimal choice for $q(z)$ is the true posterior distribution $ P(Z|Y,\theta)$ at the ML solution of $\theta$ since then $\mathcal{L}(q(Z), \theta) = \log P(Y| \theta)$. However, since we do not have this solution, we adopt an iterative procedure which involves setting $q(z) = P(Z|Y,\theta_k)$ at each step $k$. Substituting into the expression for $\mathcal{L}(q(Z), \theta)$ gives
\begin{align}
\mathcal{L}(q(Z), \theta) &= \int dZ P(Z|Y, \theta_k) \log P(Y, Z| \theta)\\
&\qquad-  \int dZ P(Z|Y, \theta_k) \log P(Y, Z| \theta_k), \nonumber \\
&=\int dZ P(Z|Y, \theta_k) \log P(Y, Z| \theta) + \mbox{const.}.
\end{align}
Thus we see that to maximise $\mathcal{L}(q(Z), \theta)$ at $\theta_k$ we need to compute the expectation value of $\log P(Y, Z|\theta)$ with respect to the posterior distribution $P(Z|Y, \theta_k)$. This is known as the E-step and it defines a function which we can subsequently maximise viz.
\begin{equation}
Q(\theta, \theta_k) = \mathbb{E}_{P(Z|Y, \theta_k)}\left[P(Y, Z| \theta)\right].
\label{optfunc1}
\end{equation}
To solve the ML problem, we now need to solve each of the following problems in order
\begin{equation}
\nabla_{\Bx} Q = 0, \quad \nabla_{\BLambda} Q = 0, \quad \mbox{and} \quad \nabla_v Q = 0.
\end{equation} 
This is known as the M-step and we can iterate between the M-step and the E-step until convergence. \\

The required joint density (also known as the complete likelihood function) for all $N$ observations $Y$ is given by the integrand of \eqref{fullmarginal} i.e.
\begin{equation}
P(Y, Z | \theta) = \prod_{i=1}^N \CN\left(\By_i|\Bmu_i(\Bx), (\tau_i \BLambda)^{-1}\right) \Gam{\tau_i| v, v}.
\label{joint}
\end{equation}
The complete log-likelihood function is therefore given by
\begin{align}
\log P(Y, Z | \theta) &\propto \sum_{i=1}^N D \log \tau_i + N \log |\BLambda| - \sum_{i=1}^N \tau_i \Delta_i^2(\Bx, \BLambda) \nonumber \\ 
 &\qquad + N v\log(v) + (v-1) \sum_{i=1}^N \log(\tau_i) - N\log(\Gamma(v)) \nonumber \\
 &\qquad - v \sum_{i=1}^N \tau_i, \label{logjoint}
\end{align}
Next, we need to compute the expectation value of $\log P(Y, Z | \theta)$ w.r.t. $P(Z|Y, \theta_k)$. Using the product rule of probability, we see that
\begin{equation}
P(Z|Y, \theta_k) = \frac{P(Z, Y|\theta_k)}{P(Y| \theta_k)} \propto P(Z, Y | \theta_k),
\end{equation}
where we have used the fact that all terms independent of $Z$ are irrelevant when computing the expectation values in \eqref{optfunc1}. We can therefore evaluate the conditional density up to a normalisation constant as
\begin{align}
P(Z | Y, \theta_k) &\propto\prod_{i=1}^N \CN\left(\By_i|\Bmu_i(\Bx_k), (\tau_i \BLambda_k)^{-1}\right) \Gam{\tau_i| v_k, v_k}, \\
&\propto \prod_{i=1}^N \tau_i^{D+v_k-1} \exp\left(-\tau_i(v_k + \Delta^2_k) \right), \\
&\propto \prod_{i=1}^N \Gam{\tau_i | v_k + D, v_k + \Delta^2_k}, \label{postZ}
\end{align}
where we have obtained the parameters of the Gamma distribution by inspection. This is actually very convenient because the terms for which we need expectation values in \eqref{optfunc1} (i.e. 1, $\tau_i$ and $\log(\tau_i)$) can all be obtained analytically using the well known properties of the Gamma distribution. They are\footnote{Note specifically that the expression for $\mathbb{E}[\log(\tau_i)]$ differs from the real valued case.}
\begin{eqnarray}
\mathbb{E}[1] &=& 1, \\
\mathbb{E}[\tau_i] &=& \frac{v_k + D}{v_k + \Delta^2(\Bx_k, \BLambda_k)}, \label{expecttau} \\
\mathbb{E}[\log(\tau_i)] &=& \psi(v_k + D) - \log(v_k + \Delta^2(\Bx_k, \BLambda_k)) \label{expectlogtau}.
\end{eqnarray}
This implies that \eqref{optfunc1} can be written as
\begin{align}
Q(\theta, \theta_k) &= D \sum_i \mathbb{E}[\log(\tau_i)] +  N \log |\BLambda| + \sum_i \Delta_i^2(\Bx, \BLambda) \mathbb{E}[\tau_i] \nonumber \\
 &\qquad+ N v\log(v) + (v-1) \sum_{i=1}^N \mathbb{E}[\log(\tau_i)] - N\log(\Gamma(v)) \nonumber \\
 &\qquad- v \sum_{i=1}^N \mathbb{E}[\tau_i].
\end{align} 
Note that the dependence on $\theta_k$ is implicit in the expressions for the expectation values. To solve the ML problem, we first need to solve 
\begin{equation}
\nabla_{\Bx} Q = 0.
\end{equation}
Since the dependence on $\Bx$ is confined to the $\Delta^2$ term, this amounts to solving
\begin{equation}
\hat{\Bx} = \underset{\Bx}{\mbox{argmin}} \sum_i (\By_i - f(\Bx))^H \BLambda (\By_i - f(\Bx)) \mathbb{E}[\tau_i].
\end{equation}
This is just a weighted NLLS problem. However, note that, the objective function is a real valued function of complex variables which is not holomorphic for all choices of $f$. Thus we require the machinery of Wirtinger calculus to tackle it. With this solution in hand, the next step is to solve
\begin{equation}
\nabla_{\BLambda} Q = 0.
\end{equation}    
It is not a fact that $\BLambda$ is diagonal; we assume this to reduce the computational cost of this step. As it stands, the Hermitian symmetry of $\BLambda$ implies that the diagonal part has to be real valued and we can proceed as normal. The solution is available in closed form and is given by
\begin{equation}
\hat{\BLambda} = \left(\frac{1}{N}\sum_{i=1}^N (\By_i - f(\hat{\Bx})(\By_i - f(\hat{\Bx})^H \mathbb{E}[\tau_i]\right)^{-1} \label{cov_eq}.
\end{equation}
Finally, we need to update the value of $v$ by solving
\begin{equation}
\nabla_{v} Q = 0 = N\log(v) + N + \sum_{i=1}^N \mathbb{E}[\log(\tau_i)] - N\psi(v) - \sum_{i=1}^N \mathbb{E}[\tau_i].
\end{equation}
This last expression needs to be solved numerically using a root finding algorithm if $v$ is continuous or using grid search if $v$ is assumed to be an integer. Once $\hat{\theta}$ has been obtained we can re-evaluate the expectation values \eqref{expecttau} and \eqref{expectlogtau} (E-step) and perform another M-step. This process is iterated until convergence and we therefore refer to it as an iteratively reweighted complex NLLS algorithm. 

\subsection*{Radio interferometric gain calibration}
\label{ssec:appendix-RI}
Let $\mathbf{d} = [d_{pq}]$ and $\mathbf{v} = [v_{pq}]$ respectively represent the vectorised observed and modelled visibilities. Using Equation \eqref{expecttau}, with $D$ replaced by the number of correlations, $n_c$, the weights are given by
\begin{equation}
w_{pq} = \frac{v+n_c}{v + (\mathbf{d}_{pq}-\mathbf{v}_{pq})^{H}{\bf{\Sigma}}^{-1}(\mathbf{d}_{pq}-\mathbf{v}_{pq})}, \label{again_w}
\end{equation}
where ${\bf{\Sigma}} = \hat{\BLambda}^{-1}$. If ${\bf{\Sigma}}$ is assumed to be $\mathbf{I}$ (identity matrix of appropriate shape), the weights become
\begin{equation}
w_{pq} = \frac{v+n_c}{v + |\mathbf{d}_{pq}-\mathbf{v}_{pq}|^{2}} \label{again_w1}.
\end{equation}

%If you want to present additional material which would interrupt the flow of the main paper,
%it can be placed in an Appendix which appears after the list of references.
%%%%%%%%%%%%%%%%%%%%%%%%%%%%%%%%%%%%%%%%%%%%%%%%%%
% Don't change these lines
\bsp    % typesetting comment
\label{lastpage}
\end{document}